\documentclass[floats,floatfix,showpacs,amssymb,prd,twocolumn,superscriptaddress,nofootinbib,nolongbibliography,reprint]{revtex4-1}

\usepackage{amssymb,amsmath,verbatim,mathtools,needspace,enumitem,etoolbox,graphicx,physics,microtype,afterpage,xspace,tabularx,lmodern,multirow}
\usepackage{gensymb}
\usepackage[normalem]{ulem}
\usepackage[dvipsnames, usenames]{xcolor}
\usepackage{xr-hyper}
\definecolor{linkcolor}{rgb}{0.0,0.3,0.5}
\usepackage[unicode, colorlinks=true, linkcolor=linkcolor, citecolor=linkcolor, filecolor=linkcolor, urlcolor=linkcolor, linktocpage, breaklinks]{hyperref}
\usepackage[all]{hypcap}
\usepackage[T1]{fontenc}
\usepackage[utf8]{inputenc}
\usepackage[usenames,dvipsnames]{xcolor}
\hypersetup{colorlinks=true,citecolor=romared,linkcolor=romared,urlcolor=romared}

\definecolor{romared}{RGB}{142,0,28}

\newcommand{\be}{\begin{equation}}
\newcommand{\ee}{\end{equation}}

\def\be{\begin{equation}}
\def\ee{\end{equation}}
\newcommand{\beq}{\begin{eqnarray}}
\newcommand{\eeq}{\end{eqnarray}}

\usepackage{makecell}
\usepackage{soul}

\newcolumntype{Y}{>{\centering\arraybackslash}X}

\makeatletter
\newcommand*{\addFileDependency}[1]{%
  \typeout{(#1)}
  \@addtofilelist{#1}
  \IfFileExists{#1}{}{\typeout{No file #1.}}
}
\makeatother

\newcommand*{\myexternaldocument}[1]{%
    \externaldocument{#1}%
    \addFileDependency{#1.tex}%
    \addFileDependency{#1.aux}%
}

\myexternaldocument{Supp}

\begin{document}

\title{Nonlinear effects in black hole ringdown}

\author{Mark Ho-Yeuk Cheung}
\affiliation{William H. Miller III Department of Physics and Astronomy, Johns Hopkins University, 3400 North Charles Street, Baltimore, Maryland 21218, USA}
\author{Vishal Baibhav}
\affiliation{Center for Interdisciplinary Exploration and Research in Astrophysics (CIERA) and Department of Physics and Astronomy,
Northwestern University, 1800 Sherman Ave, Evanston, Illinois 60201 USA}
\author{Emanuele Berti}
\affiliation{William H. Miller III Department of Physics and Astronomy, Johns Hopkins University, 3400 North Charles Street, Baltimore, Maryland, 21218, USA}
\author{Vitor Cardoso} 
\affiliation{Niels Bohr International Academy, Niels Bohr Institute, Blegdamsvej 17, 2100 Copenhagen, Denmark}
\affiliation{CENTRA, Departamento de F\'{\i}sica, Instituto Superior T\'ecnico - IST, Universidade de Lisboa -- UL, Avenida Rovisco Pais 1, 1049-001 Lisboa, Portugal}
\author{Gregorio Carullo}
\affiliation{Theoretisch-Physikalisches Institut, Friedrich-Schiller-Universit{\"a}t Jena, Fr{\"o}belstieg 1, 07743 Jena, Germany}
\affiliation{Niels Bohr International Academy, Niels Bohr Institute, Blegdamsvej 17, 2100 Copenhagen, Denmark}
\author{Roberto Cotesta}
\affiliation{William H. Miller III Department of Physics and Astronomy, Johns Hopkins University, 3400 North Charles Street, Baltimore, Maryland, 21218, USA}
\author{Walter Del Pozzo}
\affiliation{Dipartimento di Fisica “Enrico Fermi”, Università di Pisa, Pisa I-56127, Italy}
\author{Francisco Duque}
\affiliation{CENTRA, Departamento de F\'{\i}sica, Instituto Superior T\'ecnico -- IST, Universidade de Lisboa -- UL,
Avenida Rovisco Pais 1, 1049-001 Lisboa, Portugal}
\author{Thomas Helfer}
\affiliation{William H. Miller III Department of Physics and Astronomy, Johns Hopkins University, 3400 North Charles Street, Baltimore, Maryland, 21218, USA}
\author{Estuti Shukla} 
\affiliation{Institute for Gravitation and the Cosmos, Department of Physics, Pennsylvania State University, University Park, Pennsylvania 16802, USA}
\author{Kaze W. K. Wong}
\affiliation{Center for Computational Astrophysics, Flatiron Institute, New York, New York 10010, USA}

\date{\today}

\begin{abstract}
We report evidence for nonlinear modes in the ringdown stage of the gravitational waveform produced by the merger of two comparable-mass black holes. We consider both the coalescence of black hole binaries in quasicircular orbits and high-energy, head-on black hole collisions. The presence of nonlinear modes in the numerical simulations confirms that general-relativistic nonlinearities are important and must be considered in gravitational-wave data analysis. 
\end{abstract}

\maketitle
\noindent {\bf \em Introduction.} 
The birth of gravitational-wave (GW) astronomy~\cite{LIGOScientific:2016aoc} marks a new era in the exploration of strong-field gravity~\cite{Berti:2015itd,Barack:2018yly}. As the simplest macroscopic objects cloaking curvature singularities, black holes (BHs) play a special role as astrophysical laboratories to test gravity and to search for new physics~\cite{Yunes:2016jcc,Cardoso:2016ryw,Berti:2018cxi,Berti:2018vdi,Cardoso:2019rvt,LIGOScientific:2021sio}.
The structure and dynamics of BHs in our Universe is well described by the two parameters (mass $M$ and angular momentum $J$) characterizing the Kerr metric. In general relativity, the perturbed BHs formed in a binary merger approach a stationary state by emitting GWs in a discrete set of characteristic quasinormal modes (QNMs) with complex frequencies determined only by $M$ and $J$. The ``black hole spectroscopy'' program consists in observing these ``ringdown'' waves, measuring the QNM frequencies, using them to estimate mass and spin~\cite{Echeverria:1989hg}, and (if more than one mode can be observed) test that the remnant is indeed consistent with a Kerr BH~\cite{Detweiler:1980gk,Kokkotas:1999bd,Dreyer:2003bv,Berti:2005ys,Berti:2009kk}.
The observability of QNMs depends crucially on their excitation in the merger process. Even within linear perturbation theory, where one only considers linear metric perturbations to Einstein's equations in the Kerr background,
determining which modes are excited is a formidable problem~\cite{Leaver:1985ax,Leaver:1986gd,Andersson:1995zk,Berti:2006wq,Zhang:2013ksa,Hughes:2019zmt,Lim:2019xrb,Oshita:2021iyn,Lim:2022veo,London:2022urb}.

General relativity is an intrinsically nonlinear theory. The merger of two comparable-mass BHs leading to a perturbed Kerr BH is one of the most violent processes in the Universe, where these nonlinearities should play an important role. It is therefore surprising that merger simulations in numerical relativity result in a very smooth transition from inspiral to merger and ringdown~\cite{Buonanno:2006ui,Berti:2007fi}. Where are the nonlinearities of general relativity?

This state of affairs has led many (including some of us) to conjecture that nonlinear effects may be hidden behind the horizon, suppressed by the presence of a photonsphere, or even absent altogether(see e.g.~\cite{Giesler:2019uxc,Prasad:2020xgr,Okounkova:2020vwu,Jaramillo:2022mkh,Chen:2022dxt,Bhagwat:2017tkm,Cook:2020otn,Mourier:2020mwa,Finch:2021iip,Forteza:2021wfq} and references therein). In this paper we show that merger simulations of BH binaries of comparable masses in quasicircular orbits (as well as high-energy, head-on BH collisions) do, in fact, excite nonlinear modes in the ringdown stage.

\noindent {\bf \em Second-order quasinormal modes.} 
In BH perturbation theory, the GW strain and the Newman-Penrose scalar $\Psi_4$ produced by a BH merger at late times can be approximated by a linear combination of damped sinusoids (in addition to a subdominant power-law tail as well as retrograde QNMs, which we disregard here)~\cite{Leaver:1985ax,Leaver:1986gd,Berti:2005ys,Berti:2009kk},
\begin{equation}\label{eq:model}
  r h^{(1)}(t, \theta, \phi) = \sum_{n\ell m} A_{n\ell m} e^{-i (\omega_{n\ell m} t + \phi_{n\ell m})}S_{\ell m}\,, 
\end{equation}
where $r$ is the (luminosity) distance from the source.
The spin-2 spin-weighted spheroidal harmonics $S_{\ell m}=S_{\ell m}(\theta,\phi,\chi \omega_{n \ell m})$ depend on the angular variables $(\theta,\phi)$, on the complex QNM frequencies $\omega_{n\ell m}$, and on the dimensionless spin $\chi = J / M^2$ of the remnant BH~\cite{Berti:2005gp}.
This expression, found by solving the Teukolsky equation~\cite{Teukolsky:1973ha}, is valid when the GW amplitude is small enough that one can linearize Einstein's equations in the Kerr background.

At second order in the GW amplitude one finds similar equations for the second-order perturbations $h^{(2)}$, now sourced by first-order quantities~\cite{Gleiser:1995gx,Brizuela:2009qd,Ioka:2007ak,Nakano:2007cj,Pazos:2010xf,London:2014cma,Loutrel:2020wbw,Ripley:2020xby}.
Let $k$ be a generic mode, which can be either a first-order mode ($k=k_i=\ell_i m_i n_i$) or a higher-order mode. We will denote a second-order mode sourced by the first-order modes $k_1=\ell_1 m_1 n_1$ and $k_2=\ell_2 m_2 n_2$ as $k=k_1\times k_2=\ell_1 m_1 n_1 \times \ell_2 m_2 n_2$. 
From a waveform modeling point of view, the second-order modes are just damped sinusoids, like the first-order modes. 
Spin-weighted spherical harmonics, rather than spheroidal harmonics, are commonly used for waveform extraction in numerical relativity~\cite{Berti:2014fga}.
In our analysis, the index $k$ will belong to a set
$I = \{ \ell_1 m_1 n_1, \ell_2 m_2 n_2, \dots \ell_i m_i n_i \times\ell_j m_j n_j  \dots \}$ containing all the indices of the QNMs present in the $\ell m$ spin-2 spin-weighted spherical harmonic component. Then a ringdown waveform including both first- and second-order modes can be schematically written as
\begin{equation}\label{eq:model}
r h^{(2)}(t, \theta, \phi) =\sum_{\ell m} \sum_{k \in I} A_{k, \ell m} e^{-i (\omega_k t + \phi_{k, \ell m})}Y_{\ell m}\,, 
\end{equation}
where $A_{k, \ell m}$ and $\phi_{k, \ell m}$ are the amplitude and phase of the $k$th (linear or nonlinear) mode found in the $\ell m$ spin-weighted spherical harmonic component. Note that $A_{\ell_1 m n, \ell_2 m}$ could be nonzero even if $\ell_1 \neq \ell_2$ because the spheroidal harmonic $S_{\ell_1 m}$ is not necessarily orthogonal to the spherical harmonic $Y_{\ell_2 m}$, even if $\ell_1 \neq \ell_2$~\cite{Berti:2014fga}.

\begin{figure*}[t!]
	\includegraphics[width=\textwidth]{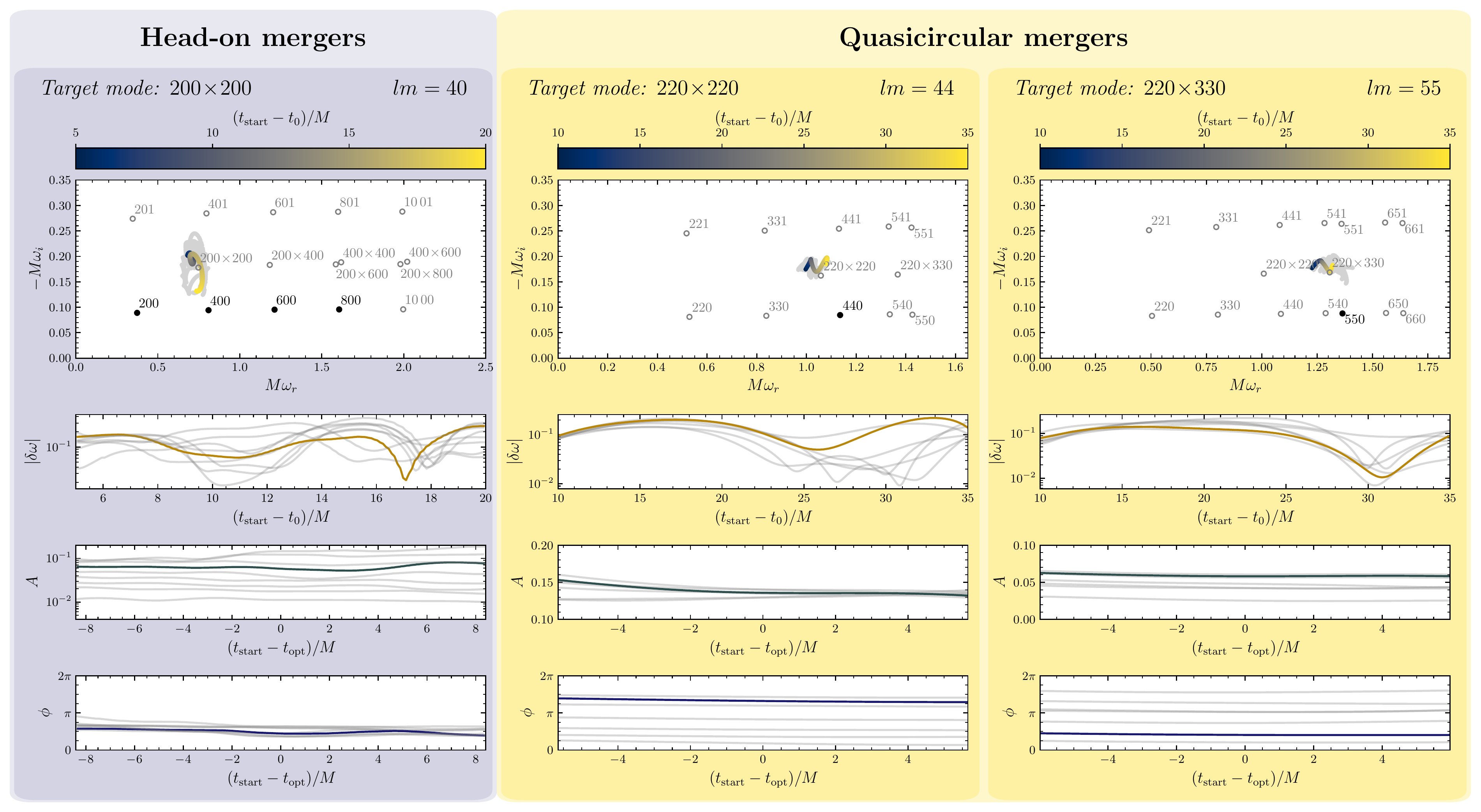}
	\caption{\label{fig:mainplot} Evidence for nonlinear effects in the ringdown.
      Left: search for the $200 \times 200$ mode in the $\ell m = 40$ multipole of ultrarelativistic head-on mergers; Center: $220 \times 220$ mode in the $\ell m = 44$ harmonic of quasicircular mergers with low mass ratio $q \leq 1.5$; right: $220 \times 330$ mode in the $\ell m = 55$ harmonic of quasicircular mergers with $1.25 \leq q \leq 2$. 
      We highlight in brighter colors the results for $\gamma = 1.5$ (left), $q = 1.22$ (the ``SXS:BBH:0305'' simulation, center) and $q = 1.88$ (``SXS:BBH:0403'' simulation, right), while we plot the results for all other simulations in gray.
      Top row: search for the second-order mode frequency.
      We use a mode with a variable complex frequency in our fitting model to search for the expected second-order modes, and we use modes with fixed frequencies (black solid circles) to remove the contribution from linear modes when they are present.
      The color scale (top bar) represents different starting times of the fit.
      For quasicircular mergers, the labeled modes correspond to those of the remnant BH in the highlighted simulation. The location of the target mode for other simulations may be slightly different, because it depends on the remnant spin.
      Second row: fractional deviation $|\delta \omega|$ of the fitted complex frequency with respect to the expected second-order mode.
      Third row: amplitude of the second-order mode when $t_{\rm start}$ is varied across a window of length $T_{0}$ centered around the value of minimum $|\delta \omega|$, $t_{\rm opt}$, and the second-order mode frequency is fixed to its expected value in the fitting model.
      Bottom row: same as the third row, but for the phase of the second-order mode.}
\end{figure*}

Because second-order QNM frequencies are sourced by first-order modes, their frequencies, amplitudes and phases are expected to obey the relationships~\cite{Ioka:2007ak,Nakano:2007cj,Pazos:2010xf,London:2014cma,Loutrel:2020wbw,Ripley:2020xby}
\begin{subequations}\label{eq:dependence}
\begin{align}
    &\omega_{k_i \times k_j} = \omega_{k_i} + \omega_{k_j} \, ,\\
    &A_{k_i \times k_j, \ell_1 m_1}
    \propto A_{k_i, \ell_2 m_2} A_{k_j, \ell_3 m_3} \, ,  \label{eq:amplitude_dependence}\\
    &\phi_{k_i \times k_j, \ell_1 m_1}
    = \phi_{k_i, \ell_2 m_2} + \phi_{k_j, \ell_3 m_3} + {\rm constant}\, . \label{eq:phase_dependence}
\end{align}
\end{subequations}

Second-order modes are a robust prediction of the perturbative expansion in general relativity. 
Other nonlinearities in the ringdown, such as the memory effect~\cite{MaganaZertuche:2021syq} or absorption-induced mode excitation~\cite{Sberna:2021eui}, have previously been observed in simulations. However nonlinear QNMs have never been confidently identified until recently~\cite{Ma:2022wpv}, with the exception of pioneering work by London {\em et al.}~\cite{London:2014cma} using greedy fitting algorithms.

\noindent {\bf \em Second-order modes in merger simulations.}
We have looked for the second-order modes in two sets of binary BH merger simulations. The first set consists of ultrarelativistic head-on collisions of equal-mass, nonspinning BHs with different boosts $\gamma$, similar to the sequences considered in Refs.~\cite{Sperhake:2008ga,Healy:2015mla}. In this one-parameter family of solutions the amplitude of the linear mode increases with the boost parameter $\gamma$, so the amplitude of the second-order modes is also a monotonic function of $\gamma$. Axial symmetry allows us to simulate this problem in two dimensions with \textsc{GRChombo}~\cite{Clough:2015sqa,Andrade:2021rbd} by applying dimensional reduction~\cite{Pretorius:2004jg,Cook:2016soy,Shibata:2010wz}, thus saving computational time and allowing for better accuracy relative to previous work~\cite{Headonpaper}. 
As the quadratic modes are sourced by a product of two first-order modes, and quadratic contributions (proportional to $Y_{\ell_i m_i} Y_{\ell_j m_j}$) overlap with $Y_{\ell_i+\ell_j m_i+m_j}$, we will look for the $\ell_i m_i n_i \times \ell_j m_j n_j$ mode in the $\ell_i+\ell_j m_i+m_j$ ringdown waveform~\cite{Nakano:2007cj,London:2014cma, Ripley:2020xby}.
For head-on collisions,  we will be fitting $r \Psi_4 = r \ddot{h}$ instead of $h$, and all of the reported amplitudes refer to $r \Psi_4$. In this case the $200$ mode dominates the ringdown of the nonspinning remnant~\cite{Sperhake:2008ga}, so we focus mainly on the $200 \times 200$ mode in the $\ell m = 40$ waveform.

The second set of simulations consists of quasicircular mergers of binary BHs with different mass ratios from the publicly available SXS waveform catalog, simulated in $(3+1)$-dimensions with the spectral code \textsc{SpEC}~\cite{Boyle:2019kee}. Recent waveforms produced using Cauchy characteristic extraction~\cite{Moxon:2020gha,Moxon:2021gbv,Mitman:2022qdl} may improve the quality of our fits, but the relatively small set of publicly available waveforms does not adequately cover the relevant parameter space for our study.
For quasicircular mergers the $220$ and $330$ modes are typically dominant (with their amplitudes depending on the mass ratio and spins of the binary), and
we focus our search on (i) the $220 \times 220$ mode in the $\ell m = 44$ waveform, and (ii) the $220 \times 330$ mode in the $\ell m = 55$ waveform.

Identifying QNMs in a waveform can be challenging, partly because of their rapid decay; for nonspinning BHs, their quality factor is of order $\sim 3$~\cite{Berti:2005ys}. The search for subdominant modes, which decay faster, requires some care. Even if their inclusion yields smaller fit residuals, consistency checks are crucial to avoid overfitting. In this work we fit the waveforms by a linear combination of damped sinusoids, as in Eq.~\eqref{eq:model}, using a least-squares fitting algorithm. The amplitude and phase of each mode are always free fitting parameters, while the complex QNM frequencies are either free or fixed depending on the mode, as shown in Fig.~\ref{fig:mainplot} and explained below.

We first try to find the second-order modes without assuming knowledge of their QNM frequencies, as follows.
We consider the QNM frequencies as free fitting parameters, and we fit the waveform with a different number of QNMs as we vary the starting time of the fit $t_{\rm start}$.
If a fitted QNM returns a frequency that is consistent with a linear mode expected to exist in the waveform over a wide range of $t_{\rm start}$, we assume that the mode is there. We then fix the frequency of that QNM (as calculated in BH perturbation theory) in our fit and we add more QNMs with free frequencies to search for additional modes. We iterate until we do not see returned frequencies that are consistent with any linear modes.
For head-on high-energy mergers, we find a combination of the modes $200, 400 \dots 10 \mskip2mu 00$ in the $\ell m = 40$ waveform due to numerical contamination between modes. 
For the SXS waveforms, we only confidently identify the $440$ mode in the $\ell m = 44$ multipole, and the $550$ mode in the $\ell m = 55$ multipole (out of all possible linear modes).
With these first-order modes identified, we use a fitting model that consists of all such modes (with {\em fixed} frequencies) to search for additional higher-order modes by adding one more damped exponential with free frequency.
As shown in the top-row panels of Fig.~\ref{fig:mainplot}, when we vary $t_{\rm start}$ relative to a reference time $t_0$ (defined to be the time of peak luminosity of the dominant $\ell m = 22$ multipole), the free mode hovers around the expected second-order mode frequency (from left to right: $\omega_{200 \times 200}$, $\omega_{220 \times 220}$, or $\omega_{220 \times 330}$, respectively). 
We do not expect the free mode frequency to converge exactly to the expected frequency due to numerical noise and contamination from other effects (such as additional nonlinearities) in the waveform, especially for modes that decay significantly faster than the dominant mode. In the Supplemental Material, we show through a controlled experiment 
that a free frequency hovering near the target mode is the expected behavior in the presence of (small) unaccounted additional modes.
We also searched for the $200 \times 400$ mode in the head-on simulations. The results (which are not as clean as those for the $200 \times 200$ mode, because $200 \times 400$ is subdominant) are shown in the Supplemental Material.
Having established the presence of nonlinear modes in the simulations, we now perform further checks to verify their physical nature.

\noindent {\bf \em Amplitude consistency check.}
We cannot exclude {\em a priori} that the new mode we found is in accidental agreement with the expected second-order QNM frequency. A nontrivial consistency test requires that, in addition to the frequency, the {\it amplitude} of the second-order modes should be consistent across different fitting ranges. To check this, we first look for the ``optimal starting time,'' $t_{\rm opt}$, for which the fractional deviation between the fitted and expected complex frequencies, i.e. $|\delta \omega| = \sqrt{\left(\frac{\omega_r - \varpi_{r}}{\varpi_{r}}\right)^2 + \left(\frac{\omega_i - \varpi_{i}}{\varpi_{i}}\right)^2}$, has a minimum. In the three cases of interest, $\varpi = \omega_{200 \times 200}$, $\omega_{220 \times 220}$ or $\omega_{220 \times 330}$, respectively.
Then we assume that the mode exists, we fix the frequency to the expected value in our fitting model, and we check the consistency of the fitted amplitude.
More explicitly, we check whether the recovered amplitude has an error smaller than 10\%  when $t_{\rm start}$ varies within a window of length $T_{0}$ centered around $t_{\rm opt}$, where $T_{0}$ is the period of oscillation of the fundamental mode across all $\ell m$ multipoles ($T_{0} = T_{200}$ for head-on mergers, and $T_{0}= T_{220}$ for inspirals).  We choose this value of $T_{0}$ because it is at least two times larger than the period of the second-order mode that we are searching for.
This threshold is further justified in the Supplemental Material by studying the impact of the numerical noise in the simulations on the quantities of interest.
Later times are excluded because the second-order mode falls below the numerical noise floor.

\begin{figure*}[t]
  \includegraphics[width=\textwidth]{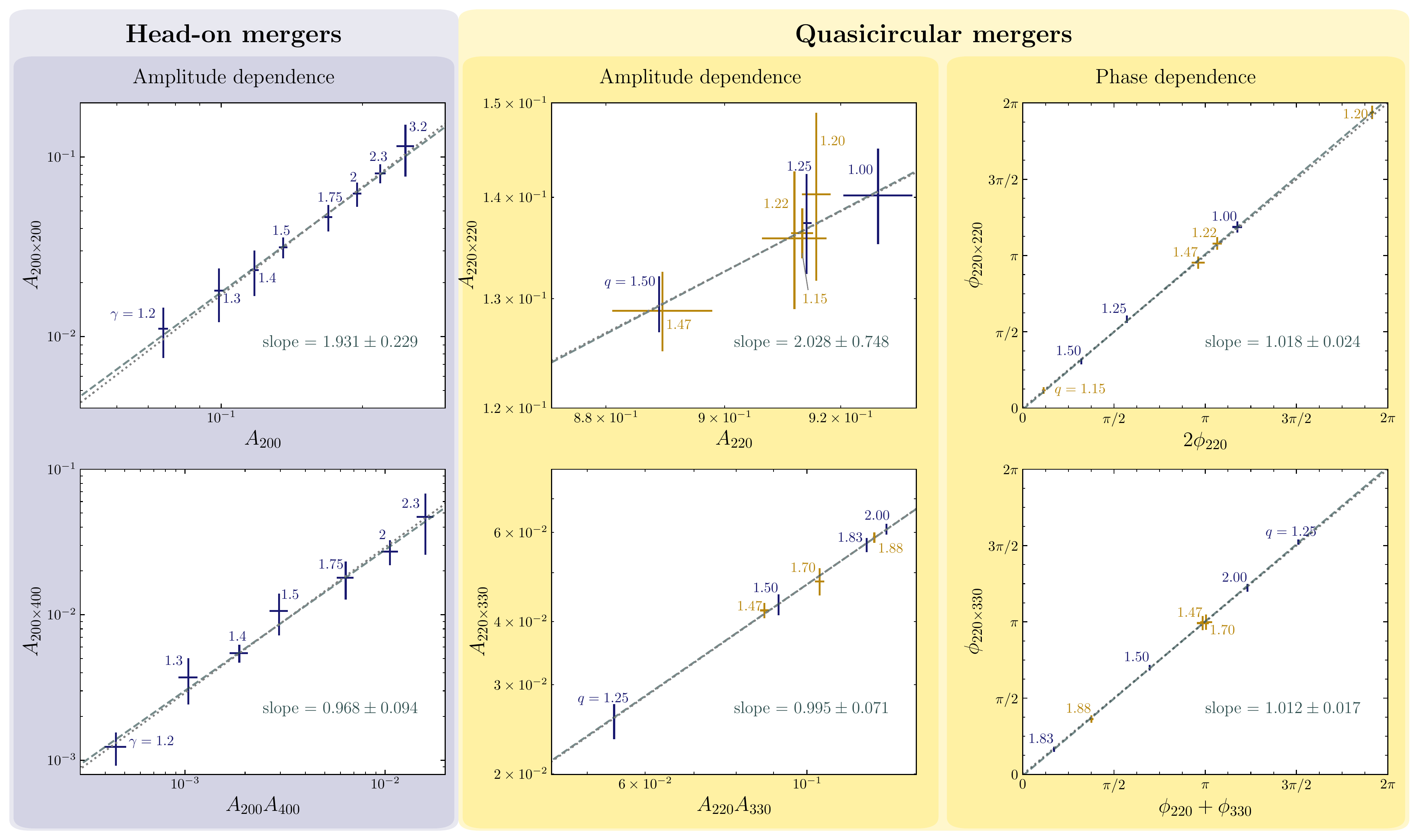}
	\caption{\label{fig:amplitude_dependence} 
      Dependence of the second-order mode amplitude (left and middle columns) and phases (right column) on the amplitudes of the first-order modes sourcing them.
      The crosses are the amplitudes or phases extracted from simulations with different boost (left column) or mass ratio (center and right columns). The width and height of the crosses correspond to the errors.
      Blue crosses represent simulations where the two BHs are initially nonspinning, while golden crosses represent those with at least one spinning BH. 
      The gray dotted line is the expected relationship between the first and second-order values with the slope fixed to either $1$ or $2$; the deep gray dashed line is a fit to the data with the slope unfixed.
      The phase dependence for head-on simulations is shown in the Supplemental Material.
}
\end{figure*}

We find that all the waveforms we considered satisfy this requirement on the amplitude. %
We also checked that the amplitudes obtained from a model with free frequency are consistent with those where the frequency is fixed, albeit with larger fluctuations, as expected.
Independently of the chosen $t_{\rm start}$, we use the convention $A_{k, \ell m} \equiv A_{k, \ell m}(t_{\rm peak})$. In other words, we take into account the known exponential time decay by extrapolating the fitted amplitudes back towards the peak of the dominant multipole.

\noindent {\bf \em Second-order amplitude dependence.}
As a more stringent check, we can verify whether the recovered second-order mode amplitudes follow the dependence predicted in Eq.~\eqref{eq:amplitude_dependence} across different simulations.
For each simulation, we extract the second-order mode amplitudes by taking the mean of the amplitude within the $T_{0}$ starting time window mentioned above.
We extract the first-order mode amplitudes after $t_{\rm start} - t_0 = 25 M$, when nonlinearities and overtones have died out.
We estimate the errors on the amplitudes as detailed in the Supplemental Material.

In Fig.~\ref{fig:amplitude_dependence} we plot the second-order mode amplitudes versus their first-order counterparts on a log-log plot. The data are consistent with a power-law dependence when the errors are taken into account.
The slope of the fitted line for $A_{200 \times 200}$ vs $A_{200}$ (in the head-on waveforms) and $A_{220 \times 220}$ vs. $A_{220}$ (in the SXS waveforms) is found to be consistent with $2$ within $1 \sigma$, as expected. 
Similarly, the slopes of the fitted lines for $A_{200 \times 400}$ vs $A_{200} A_{400}$ (for head-on waveforms) and $A_{220 \times 330}$ vs $A_{220} A_{330}$ (for SXS waveforms) are consistent with $1$.
Unsurprisingly, the $200 \times 400$ mode search results in head-on mergers are not as clean as the $200 \times 200$ results (see the Supplemental Material).

Because of numerical errors in the simulations, we can confidently identify the $220 \times 220$ mode only for SXS waveforms with mass ratio $q \le 1.5$. Since $q$ varies over a small range, the amplitudes of the $220$ mode inferred from different simulations are similar to each other, and the amplitude of the $220 \times 220$ mode does not vary much across different simulations. For this reason the data points are relatively close to each other, and the error on the slope is larger than in the other cases we considered.

\noindent {\bf \em Phase consistency.}
Similar to the amplitude tests, we can check the consistency of our fits with the fitted phases of the second-order modes. 
As shown in the bottom row of Fig.~\ref{fig:mainplot}, the fitted phases of the second-order modes vary by less than $10\% \times 2 \pi$ within the $T_0$ window.

Moreover, as the second-order modes are sourced by two linear QNMs, the relationship in Eq.~\eqref{eq:phase_dependence} between the phases of the modes should hold,
modulo (possibly) a constant phase difference that can only be computed by a Green's function calculation.
In the right column of Fig.~\ref{fig:amplitude_dependence} we show that the phases extracted from the SXS simulations follow the expected relationship.
In the Supplemental Material we show similar plots for head-on mergers. The error bars are larger, but the results are still consistent with expectations.

\noindent {\bf \em Conclusions.}
We have shown that nonlinear QNMs are excited in simulations of comparable-mass BH binary mergers in quasicircular orbits, as well as in high-energy head-on BH collisions. The detectability of nonlinear QNMs may require next-generation detectors, and it will be addressed in future work. In any case, the presence of nonlinear modes demonstrates that nonlinearities must be taken into account in the modeling of GWs from binary BH mergers, and it suggests that they may play an important role during the violent merger phase. This has far-reaching consequences for our understanding of strong-field BH dynamics and for the observational BH spectroscopy program.

\noindent {\bf \em Note added.}
While preparing this Letter, we learned that Mitman {\em et al.} conducted a similar study, whose results agree with ours~\cite{Mitman:2022qdl}.

\noindent {\bf \em Acknowledgments.} 
We thank Yanbei Chen, Macarena Lagos, Sizheng Ma, Keefe Mitman and Leo Stein for discussions.
The QNM frequencies used in this paper were computed using the \textsc{qnm} package~\cite{Stein:2019mop}.
We also made use of the \textsc{sxs} package~\cite{Boyle:2019kee} to access the SXS waveform catalog, and of the \textsc{matplotlib}~\cite{matplotlib}, \textsc{numpy}~\cite{numpy} libraries to perform our computations and create figures.
M.H.-Y.C. is a Croucher Scholar supported by the Croucher Foundation.
M.H.-Y.C., E.B., R.C. and T.H. are supported by NSF Grants No. AST-2006538, No. PHY-2207502, No. PHY-090003, and No. PHY-20043, and NASA Grants No. 19-ATP19-0051, No. 20-LPS20-0011 and No. 21-ATP21-0010. This research project was conducted using computational resources at the Maryland Advanced Research Computing Center (MARCC).
V.C.\ is a Villum Investigator and a DNRF Chair, supported by VILLUM FONDEN (Grant No.~37766) and by the Danish Research Foundation under DNRF162. V.C.\ acknowledges financial support provided under the European
Union's H2020 ERC Advanced Grant ``Black holes: gravitational engines of discovery'' Grant Agreement No.\ Gravitas–101052587. Views and opinions expressed are however those of the author only and do not necessarily reflect those of the European Union or the European Research Council. Neither the European Union nor the granting authority can be held responsible for them.
G.C. acknowledges support by the Della Riccia Foundation under an Early Career Scientist Fellowship, and funding from the European Union’s Horizon 2020 research and innovation program under the Marie Sklodowska-Curie Grant Agreement No. 847523 ‘INTERACTIONS’, from the Villum Investigator program supported by the VILLUM Foundation (Grant No. VIL37766) and the DNRF Chair program (Grant No. DNRF162) by the Danish National Research Foundation.
K.W.K.W. is supported by the Simons Foundation.
This project has received funding from the European Union's Horizon 2020 research and innovation programme under the Marie Sklodowska-Curie Grant Agreement No. 101007855.
F.D. acknowledges financial support provided by FCT/Portugal through Grant No. SFRH/BD/143657/2019. 
We thank FCT for financial support through Project~No.~UIDB/00099/2020.
We acknowledge financial support provided by FCT/Portugal through Grants No. PTDC/MAT-APL/30043/2017 and No. PTDC/FIS-AST/7002/2020.
Part of E.B.’s work was performed at the Aspen Center for Physics, which is supported by National Science Foundation grant PHY-1607611.
The authors acknowledge the Texas Advanced Computing Center (TACC) at The University of Texas at Austin for providing {HPC, visualization, database, or grid} resources that have contributed to the research results reported within this paper \cite{10.1145/3311790.3396656}.\footnote{\url{http://www.tacc.utexas.edu}}

\bibliography{SecondOrderQNM}

\begin{thebibliography}{67}%
\makeatletter
\providecommand \@ifxundefined [1]{%
 \@ifx{#1\undefined}
}%
\providecommand \@ifnum [1]{%
 \ifnum #1\expandafter \@firstoftwo
 \else \expandafter \@secondoftwo
 \fi
}%
\providecommand \@ifx [1]{%
 \ifx #1\expandafter \@firstoftwo
 \else \expandafter \@secondoftwo
 \fi
}%
\providecommand \natexlab [1]{#1}%
\providecommand \enquote  [1]{``#1''}%
\providecommand \bibnamefont  [1]{#1}%
\providecommand \bibfnamefont [1]{#1}%
\providecommand \citenamefont [1]{#1}%
\providecommand \href@noop [0]{\@secondoftwo}%
\providecommand \href [0]{\begingroup \@sanitize@url \@href}%
\providecommand \@href[1]{\@@startlink{#1}\@@href}%
\providecommand \@@href[1]{\endgroup#1\@@endlink}%
\providecommand \@sanitize@url [0]{\catcode `\\12\catcode `\$12\catcode
  `\&12\catcode `\#12\catcode `\^12\catcode `\_12\catcode `\%12\relax}%
\providecommand \@@startlink[1]{}%
\providecommand \@@endlink[0]{}%
\providecommand \url  [0]{\begingroup\@sanitize@url \@url }%
\providecommand \@url [1]{\endgroup\@href {#1}{\urlprefix }}%
\providecommand \urlprefix  [0]{URL }%
\providecommand \Eprint [0]{\href }%
\providecommand \doibase [0]{http://dx.doi.org/}%
\providecommand \selectlanguage [0]{\@gobble}%
\providecommand \bibinfo  [0]{\@secondoftwo}%
\providecommand \bibfield  [0]{\@secondoftwo}%
\providecommand \translation [1]{[#1]}%
\providecommand \BibitemOpen [0]{}%
\providecommand \bibitemStop [0]{}%
\providecommand \bibitemNoStop [0]{.\EOS\space}%
\providecommand \EOS [0]{\spacefactor3000\relax}%
\providecommand \BibitemShut  [1]{\csname bibitem#1\endcsname}%
\let\auto@bib@innerbib\@empty
\bibitem [{\citenamefont {Abbott}\ \emph {et~al.}(2016)\citenamefont {Abbott}
  \emph {et~al.}}]{LIGOScientific:2016aoc}%
  \BibitemOpen
  \bibfield  {author} {\bibinfo {author} {\bibfnamefont {B.~P.}\ \bibnamefont
  {Abbott}} \emph {et~al.} (\bibinfo {collaboration} {LIGO Scientific,
  Virgo}),\ }\href {\doibase 10.1103/PhysRevLett.116.061102} {\bibfield
  {journal} {\bibinfo  {journal} {Phys. Rev. Lett.}\ }\textbf {\bibinfo
  {volume} {116}},\ \bibinfo {pages} {061102} (\bibinfo {year} {2016})},\
  \Eprint {http://arxiv.org/abs/1602.03837} {arXiv:1602.03837 [gr-qc]}
  \BibitemShut {NoStop}%
\bibitem [{\citenamefont {Berti}\ \emph {et~al.}(2015)\citenamefont {Berti}
  \emph {et~al.}}]{Berti:2015itd}%
  \BibitemOpen
  \bibfield  {author} {\bibinfo {author} {\bibfnamefont {E.}~\bibnamefont
  {Berti}} \emph {et~al.},\ }\href {\doibase 10.1088/0264-9381/32/24/243001}
  {\bibfield  {journal} {\bibinfo  {journal} {Class. Quant. Grav.}\ }\textbf
  {\bibinfo {volume} {32}},\ \bibinfo {pages} {243001} (\bibinfo {year}
  {2015})},\ \Eprint {http://arxiv.org/abs/1501.07274} {arXiv:1501.07274
  [gr-qc]} \BibitemShut {NoStop}%
\bibitem [{\citenamefont {Barack}\ \emph {et~al.}(2019)\citenamefont {Barack}
  \emph {et~al.}}]{Barack:2018yly}%
  \BibitemOpen
  \bibfield  {author} {\bibinfo {author} {\bibfnamefont {L.}~\bibnamefont
  {Barack}} \emph {et~al.},\ }\href {\doibase 10.1088/1361-6382/ab0587}
  {\bibfield  {journal} {\bibinfo  {journal} {Class. Quant. Grav.}\ }\textbf
  {\bibinfo {volume} {36}},\ \bibinfo {pages} {143001} (\bibinfo {year}
  {2019})},\ \Eprint {http://arxiv.org/abs/1806.05195} {arXiv:1806.05195
  [gr-qc]} \BibitemShut {NoStop}%
\bibitem [{\citenamefont {Yunes}\ \emph {et~al.}(2016)\citenamefont {Yunes},
  \citenamefont {Yagi},\ and\ \citenamefont {Pretorius}}]{Yunes:2016jcc}%
  \BibitemOpen
  \bibfield  {author} {\bibinfo {author} {\bibfnamefont {N.}~\bibnamefont
  {Yunes}}, \bibinfo {author} {\bibfnamefont {K.}~\bibnamefont {Yagi}}, \ and\
  \bibinfo {author} {\bibfnamefont {F.}~\bibnamefont {Pretorius}},\ }\href
  {\doibase 10.1103/PhysRevD.94.084002} {\bibfield  {journal} {\bibinfo
  {journal} {Phys. Rev. D}\ }\textbf {\bibinfo {volume} {94}},\ \bibinfo
  {pages} {084002} (\bibinfo {year} {2016})},\ \Eprint
  {http://arxiv.org/abs/1603.08955} {arXiv:1603.08955 [gr-qc]} \BibitemShut
  {NoStop}%
\bibitem [{\citenamefont {Cardoso}\ and\ \citenamefont
  {Gualtieri}(2016)}]{Cardoso:2016ryw}%
  \BibitemOpen
  \bibfield  {author} {\bibinfo {author} {\bibfnamefont {V.}~\bibnamefont
  {Cardoso}}\ and\ \bibinfo {author} {\bibfnamefont {L.}~\bibnamefont
  {Gualtieri}},\ }\href {\doibase 10.1088/0264-9381/33/17/174001} {\bibfield
  {journal} {\bibinfo  {journal} {Class. Quant. Grav.}\ }\textbf {\bibinfo
  {volume} {33}},\ \bibinfo {pages} {174001} (\bibinfo {year} {2016})},\
  \Eprint {http://arxiv.org/abs/1607.03133} {arXiv:1607.03133 [gr-qc]}
  \BibitemShut {NoStop}%
\bibitem [{\citenamefont {Berti}\ \emph
  {et~al.}(2018{\natexlab{a}})\citenamefont {Berti}, \citenamefont {Yagi},\
  and\ \citenamefont {Yunes}}]{Berti:2018cxi}%
  \BibitemOpen
  \bibfield  {author} {\bibinfo {author} {\bibfnamefont {E.}~\bibnamefont
  {Berti}}, \bibinfo {author} {\bibfnamefont {K.}~\bibnamefont {Yagi}}, \ and\
  \bibinfo {author} {\bibfnamefont {N.}~\bibnamefont {Yunes}},\ }\href
  {\doibase 10.1007/s10714-018-2362-8} {\bibfield  {journal} {\bibinfo
  {journal} {Gen. Rel. Grav.}\ }\textbf {\bibinfo {volume} {50}},\ \bibinfo
  {pages} {46} (\bibinfo {year} {2018}{\natexlab{a}})},\ \Eprint
  {http://arxiv.org/abs/1801.03208} {arXiv:1801.03208 [gr-qc]} \BibitemShut
  {NoStop}%
\bibitem [{\citenamefont {Berti}\ \emph
  {et~al.}(2018{\natexlab{b}})\citenamefont {Berti}, \citenamefont {Yagi},
  \citenamefont {Yang},\ and\ \citenamefont {Yunes}}]{Berti:2018vdi}%
  \BibitemOpen
  \bibfield  {author} {\bibinfo {author} {\bibfnamefont {E.}~\bibnamefont
  {Berti}}, \bibinfo {author} {\bibfnamefont {K.}~\bibnamefont {Yagi}},
  \bibinfo {author} {\bibfnamefont {H.}~\bibnamefont {Yang}}, \ and\ \bibinfo
  {author} {\bibfnamefont {N.}~\bibnamefont {Yunes}},\ }\href {\doibase
  10.1007/s10714-018-2372-6} {\bibfield  {journal} {\bibinfo  {journal} {Gen.
  Rel. Grav.}\ }\textbf {\bibinfo {volume} {50}},\ \bibinfo {pages} {49}
  (\bibinfo {year} {2018}{\natexlab{b}})},\ \Eprint
  {http://arxiv.org/abs/1801.03587} {arXiv:1801.03587 [gr-qc]} \BibitemShut
  {NoStop}%
\bibitem [{\citenamefont {Cardoso}\ and\ \citenamefont
  {Pani}(2019)}]{Cardoso:2019rvt}%
  \BibitemOpen
  \bibfield  {author} {\bibinfo {author} {\bibfnamefont {V.}~\bibnamefont
  {Cardoso}}\ and\ \bibinfo {author} {\bibfnamefont {P.}~\bibnamefont {Pani}},\
  }\href {\doibase 10.1007/s41114-019-0020-4} {\bibfield  {journal} {\bibinfo
  {journal} {Living Rev. Rel.}\ }\textbf {\bibinfo {volume} {22}},\ \bibinfo
  {pages} {4} (\bibinfo {year} {2019})},\ \Eprint
  {http://arxiv.org/abs/1904.05363} {arXiv:1904.05363 [gr-qc]} \BibitemShut
  {NoStop}%
\bibitem [{\citenamefont {Abbott}\ \emph {et~al.}(2021)\citenamefont {Abbott}
  \emph {et~al.}}]{LIGOScientific:2021sio}%
  \BibitemOpen
  \bibfield  {author} {\bibinfo {author} {\bibfnamefont {R.}~\bibnamefont
  {Abbott}} \emph {et~al.} (\bibinfo {collaboration} {LIGO Scientific, VIRGO,
  KAGRA}),\ }\href@noop {} {\  (\bibinfo {year} {2021})},\ \Eprint
  {http://arxiv.org/abs/2112.06861} {arXiv:2112.06861 [gr-qc]} \BibitemShut
  {NoStop}%
\bibitem [{\citenamefont {Echeverria}(1989)}]{Echeverria:1989hg}%
  \BibitemOpen
  \bibfield  {author} {\bibinfo {author} {\bibfnamefont {F.}~\bibnamefont
  {Echeverria}},\ }\href {\doibase 10.1103/PhysRevD.40.3194} {\bibfield
  {journal} {\bibinfo  {journal} {Phys. Rev. D}\ }\textbf {\bibinfo {volume}
  {40}},\ \bibinfo {pages} {3194} (\bibinfo {year} {1989})}\BibitemShut
  {NoStop}%
\bibitem [{\citenamefont {Detweiler}(1980)}]{Detweiler:1980gk}%
  \BibitemOpen
  \bibfield  {author} {\bibinfo {author} {\bibfnamefont {S.~L.}\ \bibnamefont
  {Detweiler}},\ }\href {\doibase 10.1086/158109} {\bibfield  {journal}
  {\bibinfo  {journal} {Astrophys. J.}\ }\textbf {\bibinfo {volume} {239}},\
  \bibinfo {pages} {292} (\bibinfo {year} {1980})}\BibitemShut {NoStop}%
\bibitem [{\citenamefont {Kokkotas}\ and\ \citenamefont
  {Schmidt}(1999)}]{Kokkotas:1999bd}%
  \BibitemOpen
  \bibfield  {author} {\bibinfo {author} {\bibfnamefont {K.~D.}\ \bibnamefont
  {Kokkotas}}\ and\ \bibinfo {author} {\bibfnamefont {B.~G.}\ \bibnamefont
  {Schmidt}},\ }\href {\doibase 10.12942/lrr-1999-2} {\bibfield  {journal}
  {\bibinfo  {journal} {Living Rev. Rel.}\ }\textbf {\bibinfo {volume} {2}},\
  \bibinfo {pages} {2} (\bibinfo {year} {1999})},\ \Eprint
  {http://arxiv.org/abs/gr-qc/9909058} {arXiv:gr-qc/9909058} \BibitemShut
  {NoStop}%
\bibitem [{\citenamefont {Dreyer}\ \emph {et~al.}(2004)\citenamefont {Dreyer},
  \citenamefont {Kelly}, \citenamefont {Krishnan}, \citenamefont {Finn},
  \citenamefont {Garrison},\ and\ \citenamefont
  {Lopez-Aleman}}]{Dreyer:2003bv}%
  \BibitemOpen
  \bibfield  {author} {\bibinfo {author} {\bibfnamefont {O.}~\bibnamefont
  {Dreyer}}, \bibinfo {author} {\bibfnamefont {B.~J.}\ \bibnamefont {Kelly}},
  \bibinfo {author} {\bibfnamefont {B.}~\bibnamefont {Krishnan}}, \bibinfo
  {author} {\bibfnamefont {L.~S.}\ \bibnamefont {Finn}}, \bibinfo {author}
  {\bibfnamefont {D.}~\bibnamefont {Garrison}}, \ and\ \bibinfo {author}
  {\bibfnamefont {R.}~\bibnamefont {Lopez-Aleman}},\ }\href {\doibase
  10.1088/0264-9381/21/4/003} {\bibfield  {journal} {\bibinfo  {journal}
  {Class. Quant. Grav.}\ }\textbf {\bibinfo {volume} {21}},\ \bibinfo {pages}
  {787} (\bibinfo {year} {2004})},\ \Eprint
  {http://arxiv.org/abs/gr-qc/0309007} {arXiv:gr-qc/0309007} \BibitemShut
  {NoStop}%
\bibitem [{\citenamefont {Berti}\ \emph
  {et~al.}(2006{\natexlab{a}})\citenamefont {Berti}, \citenamefont {Cardoso},\
  and\ \citenamefont {Will}}]{Berti:2005ys}%
  \BibitemOpen
  \bibfield  {author} {\bibinfo {author} {\bibfnamefont {E.}~\bibnamefont
  {Berti}}, \bibinfo {author} {\bibfnamefont {V.}~\bibnamefont {Cardoso}}, \
  and\ \bibinfo {author} {\bibfnamefont {C.~M.}\ \bibnamefont {Will}},\ }\href
  {\doibase 10.1103/PhysRevD.73.064030} {\bibfield  {journal} {\bibinfo
  {journal} {Phys. Rev. D}\ }\textbf {\bibinfo {volume} {73}},\ \bibinfo
  {pages} {064030} (\bibinfo {year} {2006}{\natexlab{a}})},\ \Eprint
  {http://arxiv.org/abs/gr-qc/0512160} {arXiv:gr-qc/0512160} \BibitemShut
  {NoStop}%
\bibitem [{\citenamefont {Berti}\ \emph {et~al.}(2009)\citenamefont {Berti},
  \citenamefont {Cardoso},\ and\ \citenamefont {Starinets}}]{Berti:2009kk}%
  \BibitemOpen
  \bibfield  {author} {\bibinfo {author} {\bibfnamefont {E.}~\bibnamefont
  {Berti}}, \bibinfo {author} {\bibfnamefont {V.}~\bibnamefont {Cardoso}}, \
  and\ \bibinfo {author} {\bibfnamefont {A.~O.}\ \bibnamefont {Starinets}},\
  }\href {\doibase 10.1088/0264-9381/26/16/163001} {\bibfield  {journal}
  {\bibinfo  {journal} {Class. Quant. Grav.}\ }\textbf {\bibinfo {volume}
  {26}},\ \bibinfo {pages} {163001} (\bibinfo {year} {2009})},\ \Eprint
  {http://arxiv.org/abs/0905.2975} {arXiv:0905.2975 [gr-qc]} \BibitemShut
  {NoStop}%
\bibitem [{\citenamefont {Leaver}(1985)}]{Leaver:1985ax}%
  \BibitemOpen
  \bibfield  {author} {\bibinfo {author} {\bibfnamefont {E.~W.}\ \bibnamefont
  {Leaver}},\ }\href {\doibase 10.1098/rspa.1985.0119} {\bibfield  {journal}
  {\bibinfo  {journal} {Proc. Roy. Soc. Lond. A}\ }\textbf {\bibinfo {volume}
  {402}},\ \bibinfo {pages} {285} (\bibinfo {year} {1985})}\BibitemShut
  {NoStop}%
\bibitem [{\citenamefont {Leaver}(1986)}]{Leaver:1986gd}%
  \BibitemOpen
  \bibfield  {author} {\bibinfo {author} {\bibfnamefont {E.~W.}\ \bibnamefont
  {Leaver}},\ }\href {\doibase 10.1103/PhysRevD.34.384} {\bibfield  {journal}
  {\bibinfo  {journal} {Phys. Rev. D}\ }\textbf {\bibinfo {volume} {34}},\
  \bibinfo {pages} {384} (\bibinfo {year} {1986})}\BibitemShut {NoStop}%
\bibitem [{\citenamefont {Andersson}(1995)}]{Andersson:1995zk}%
  \BibitemOpen
  \bibfield  {author} {\bibinfo {author} {\bibfnamefont {N.}~\bibnamefont
  {Andersson}},\ }\href {\doibase 10.1103/PhysRevD.51.353} {\bibfield
  {journal} {\bibinfo  {journal} {Phys. Rev. D}\ }\textbf {\bibinfo {volume}
  {51}},\ \bibinfo {pages} {353} (\bibinfo {year} {1995})}\BibitemShut
  {NoStop}%
\bibitem [{\citenamefont {Berti}\ and\ \citenamefont
  {Cardoso}(2006)}]{Berti:2006wq}%
  \BibitemOpen
  \bibfield  {author} {\bibinfo {author} {\bibfnamefont {E.}~\bibnamefont
  {Berti}}\ and\ \bibinfo {author} {\bibfnamefont {V.}~\bibnamefont
  {Cardoso}},\ }\href {\doibase 10.1103/PhysRevD.74.104020} {\bibfield
  {journal} {\bibinfo  {journal} {Phys. Rev. D}\ }\textbf {\bibinfo {volume}
  {74}},\ \bibinfo {pages} {104020} (\bibinfo {year} {2006})},\ \Eprint
  {http://arxiv.org/abs/gr-qc/0605118} {arXiv:gr-qc/0605118} \BibitemShut
  {NoStop}%
\bibitem [{\citenamefont {Zhang}\ \emph {et~al.}(2013)\citenamefont {Zhang},
  \citenamefont {Berti},\ and\ \citenamefont {Cardoso}}]{Zhang:2013ksa}%
  \BibitemOpen
  \bibfield  {author} {\bibinfo {author} {\bibfnamefont {Z.}~\bibnamefont
  {Zhang}}, \bibinfo {author} {\bibfnamefont {E.}~\bibnamefont {Berti}}, \ and\
  \bibinfo {author} {\bibfnamefont {V.}~\bibnamefont {Cardoso}},\ }\href
  {\doibase 10.1103/PhysRevD.88.044018} {\bibfield  {journal} {\bibinfo
  {journal} {Phys. Rev. D}\ }\textbf {\bibinfo {volume} {88}},\ \bibinfo
  {pages} {044018} (\bibinfo {year} {2013})},\ \Eprint
  {http://arxiv.org/abs/1305.4306} {arXiv:1305.4306 [gr-qc]} \BibitemShut
  {NoStop}%
\bibitem [{\citenamefont {Hughes}\ \emph {et~al.}(2019)\citenamefont {Hughes},
  \citenamefont {Apte}, \citenamefont {Khanna},\ and\ \citenamefont
  {Lim}}]{Hughes:2019zmt}%
  \BibitemOpen
  \bibfield  {author} {\bibinfo {author} {\bibfnamefont {S.~A.}\ \bibnamefont
  {Hughes}}, \bibinfo {author} {\bibfnamefont {A.}~\bibnamefont {Apte}},
  \bibinfo {author} {\bibfnamefont {G.}~\bibnamefont {Khanna}}, \ and\ \bibinfo
  {author} {\bibfnamefont {H.}~\bibnamefont {Lim}},\ }\href {\doibase
  10.1103/PhysRevLett.123.161101} {\bibfield  {journal} {\bibinfo  {journal}
  {Phys. Rev. Lett.}\ }\textbf {\bibinfo {volume} {123}},\ \bibinfo {pages}
  {161101} (\bibinfo {year} {2019})},\ \Eprint
  {http://arxiv.org/abs/1901.05900} {arXiv:1901.05900 [gr-qc]} \BibitemShut
  {NoStop}%
\bibitem [{\citenamefont {Lim}\ \emph {et~al.}(2019)\citenamefont {Lim},
  \citenamefont {Khanna}, \citenamefont {Apte},\ and\ \citenamefont
  {Hughes}}]{Lim:2019xrb}%
  \BibitemOpen
  \bibfield  {author} {\bibinfo {author} {\bibfnamefont {H.}~\bibnamefont
  {Lim}}, \bibinfo {author} {\bibfnamefont {G.}~\bibnamefont {Khanna}},
  \bibinfo {author} {\bibfnamefont {A.}~\bibnamefont {Apte}}, \ and\ \bibinfo
  {author} {\bibfnamefont {S.~A.}\ \bibnamefont {Hughes}},\ }\href {\doibase
  10.1103/PhysRevD.100.084032} {\bibfield  {journal} {\bibinfo  {journal}
  {Phys. Rev. D}\ }\textbf {\bibinfo {volume} {100}},\ \bibinfo {pages}
  {084032} (\bibinfo {year} {2019})},\ \Eprint
  {http://arxiv.org/abs/1901.05902} {arXiv:1901.05902 [gr-qc]} \BibitemShut
  {NoStop}%
\bibitem [{\citenamefont {Oshita}(2021)}]{Oshita:2021iyn}%
  \BibitemOpen
  \bibfield  {author} {\bibinfo {author} {\bibfnamefont {N.}~\bibnamefont
  {Oshita}},\ }\href {\doibase 10.1103/PhysRevD.104.124032} {\bibfield
  {journal} {\bibinfo  {journal} {Phys. Rev. D}\ }\textbf {\bibinfo {volume}
  {104}},\ \bibinfo {pages} {124032} (\bibinfo {year} {2021})},\ \Eprint
  {http://arxiv.org/abs/2109.09757} {arXiv:2109.09757 [gr-qc]} \BibitemShut
  {NoStop}%
\bibitem [{\citenamefont {Lim}\ \emph {et~al.}(2022)\citenamefont {Lim},
  \citenamefont {Khanna},\ and\ \citenamefont {Hughes}}]{Lim:2022veo}%
  \BibitemOpen
  \bibfield  {author} {\bibinfo {author} {\bibfnamefont {H.}~\bibnamefont
  {Lim}}, \bibinfo {author} {\bibfnamefont {G.}~\bibnamefont {Khanna}}, \ and\
  \bibinfo {author} {\bibfnamefont {S.~A.}\ \bibnamefont {Hughes}},\ }\href
  {\doibase 10.1103/PhysRevD.105.124030} {\bibfield  {journal} {\bibinfo
  {journal} {Phys. Rev. D}\ }\textbf {\bibinfo {volume} {105}},\ \bibinfo
  {pages} {124030} (\bibinfo {year} {2022})},\ \Eprint
  {http://arxiv.org/abs/2204.06007} {arXiv:2204.06007 [gr-qc]} \BibitemShut
  {NoStop}%
\bibitem [{\citenamefont {London}\ and\ \citenamefont
  {Hughes}(2022)}]{London:2022urb}%
  \BibitemOpen
  \bibfield  {author} {\bibinfo {author} {\bibfnamefont {L.}~\bibnamefont
  {London}}\ and\ \bibinfo {author} {\bibfnamefont {S.~A.}\ \bibnamefont
  {Hughes}},\ }\href@noop {} {\  (\bibinfo {year} {2022})},\ \Eprint
  {http://arxiv.org/abs/2206.15246} {arXiv:2206.15246 [gr-qc]} \BibitemShut
  {NoStop}%
\bibitem [{\citenamefont {Buonanno}\ \emph {et~al.}(2007)\citenamefont
  {Buonanno}, \citenamefont {Cook},\ and\ \citenamefont
  {Pretorius}}]{Buonanno:2006ui}%
  \BibitemOpen
  \bibfield  {author} {\bibinfo {author} {\bibfnamefont {A.}~\bibnamefont
  {Buonanno}}, \bibinfo {author} {\bibfnamefont {G.~B.}\ \bibnamefont {Cook}},
  \ and\ \bibinfo {author} {\bibfnamefont {F.}~\bibnamefont {Pretorius}},\
  }\href {\doibase 10.1103/PhysRevD.75.124018} {\bibfield  {journal} {\bibinfo
  {journal} {Phys. Rev. D}\ }\textbf {\bibinfo {volume} {75}},\ \bibinfo
  {pages} {124018} (\bibinfo {year} {2007})},\ \Eprint
  {http://arxiv.org/abs/gr-qc/0610122} {arXiv:gr-qc/0610122} \BibitemShut
  {NoStop}%
\bibitem [{\citenamefont {Berti}\ \emph {et~al.}(2007)\citenamefont {Berti},
  \citenamefont {Cardoso}, \citenamefont {Gonzalez}, \citenamefont {Sperhake},
  \citenamefont {Hannam}, \citenamefont {Husa},\ and\ \citenamefont
  {Bruegmann}}]{Berti:2007fi}%
  \BibitemOpen
  \bibfield  {author} {\bibinfo {author} {\bibfnamefont {E.}~\bibnamefont
  {Berti}}, \bibinfo {author} {\bibfnamefont {V.}~\bibnamefont {Cardoso}},
  \bibinfo {author} {\bibfnamefont {J.~A.}\ \bibnamefont {Gonzalez}}, \bibinfo
  {author} {\bibfnamefont {U.}~\bibnamefont {Sperhake}}, \bibinfo {author}
  {\bibfnamefont {M.}~\bibnamefont {Hannam}}, \bibinfo {author} {\bibfnamefont
  {S.}~\bibnamefont {Husa}}, \ and\ \bibinfo {author} {\bibfnamefont
  {B.}~\bibnamefont {Bruegmann}},\ }\href {\doibase 10.1103/PhysRevD.76.064034}
  {\bibfield  {journal} {\bibinfo  {journal} {Phys. Rev. D}\ }\textbf {\bibinfo
  {volume} {76}},\ \bibinfo {pages} {064034} (\bibinfo {year} {2007})},\
  \Eprint {http://arxiv.org/abs/gr-qc/0703053} {arXiv:gr-qc/0703053}
  \BibitemShut {NoStop}%
\bibitem [{\citenamefont {Giesler}\ \emph {et~al.}(2019)\citenamefont
  {Giesler}, \citenamefont {Isi}, \citenamefont {Scheel},\ and\ \citenamefont
  {Teukolsky}}]{Giesler:2019uxc}%
  \BibitemOpen
  \bibfield  {author} {\bibinfo {author} {\bibfnamefont {M.}~\bibnamefont
  {Giesler}}, \bibinfo {author} {\bibfnamefont {M.}~\bibnamefont {Isi}},
  \bibinfo {author} {\bibfnamefont {M.~A.}\ \bibnamefont {Scheel}}, \ and\
  \bibinfo {author} {\bibfnamefont {S.}~\bibnamefont {Teukolsky}},\ }\href
  {\doibase 10.1103/PhysRevX.9.041060} {\bibfield  {journal} {\bibinfo
  {journal} {Phys. Rev. X}\ }\textbf {\bibinfo {volume} {9}},\ \bibinfo {pages}
  {041060} (\bibinfo {year} {2019})},\ \Eprint
  {http://arxiv.org/abs/1903.08284} {arXiv:1903.08284 [gr-qc]} \BibitemShut
  {NoStop}%
\bibitem [{\citenamefont {Prasad}\ \emph {et~al.}(2020)\citenamefont {Prasad},
  \citenamefont {Gupta}, \citenamefont {Bose}, \citenamefont {Krishnan},\ and\
  \citenamefont {Schnetter}}]{Prasad:2020xgr}%
  \BibitemOpen
  \bibfield  {author} {\bibinfo {author} {\bibfnamefont {V.}~\bibnamefont
  {Prasad}}, \bibinfo {author} {\bibfnamefont {A.}~\bibnamefont {Gupta}},
  \bibinfo {author} {\bibfnamefont {S.}~\bibnamefont {Bose}}, \bibinfo {author}
  {\bibfnamefont {B.}~\bibnamefont {Krishnan}}, \ and\ \bibinfo {author}
  {\bibfnamefont {E.}~\bibnamefont {Schnetter}},\ }\href {\doibase
  10.1103/PhysRevLett.125.121101} {\bibfield  {journal} {\bibinfo  {journal}
  {Phys. Rev. Lett.}\ }\textbf {\bibinfo {volume} {125}},\ \bibinfo {pages}
  {121101} (\bibinfo {year} {2020})},\ \Eprint
  {http://arxiv.org/abs/2003.06215} {arXiv:2003.06215 [gr-qc]} \BibitemShut
  {NoStop}%
\bibitem [{\citenamefont {Okounkova}(2020)}]{Okounkova:2020vwu}%
  \BibitemOpen
  \bibfield  {author} {\bibinfo {author} {\bibfnamefont {M.}~\bibnamefont
  {Okounkova}},\ }\href@noop {} {\  (\bibinfo {year} {2020})},\ \Eprint
  {http://arxiv.org/abs/2004.00671} {arXiv:2004.00671 [gr-qc]} \BibitemShut
  {NoStop}%
\bibitem [{\citenamefont {Jaramillo}\ and\ \citenamefont
  {Krishnan}(2022)}]{Jaramillo:2022mkh}%
  \BibitemOpen
  \bibfield  {author} {\bibinfo {author} {\bibfnamefont {J.~L.}\ \bibnamefont
  {Jaramillo}}\ and\ \bibinfo {author} {\bibfnamefont {B.}~\bibnamefont
  {Krishnan}},\ }\href@noop {} {\  (\bibinfo {year} {2022})},\ \Eprint
  {http://arxiv.org/abs/2206.02117} {arXiv:2206.02117 [gr-qc]} \BibitemShut
  {NoStop}%
\bibitem [{\citenamefont {Chen}\ \emph {et~al.}(2022)\citenamefont {Chen} \emph
  {et~al.}}]{Chen:2022dxt}%
  \BibitemOpen
  \bibfield  {author} {\bibinfo {author} {\bibfnamefont {Y.}~\bibnamefont
  {Chen}} \emph {et~al.},\ }\href@noop {} {\  (\bibinfo {year} {2022})},\
  \Eprint {http://arxiv.org/abs/2208.02965} {arXiv:2208.02965 [gr-qc]}
  \BibitemShut {NoStop}%
\bibitem [{\citenamefont {Bhagwat}\ \emph {et~al.}(2018)\citenamefont
  {Bhagwat}, \citenamefont {Okounkova}, \citenamefont {Ballmer}, \citenamefont
  {Brown}, \citenamefont {Giesler}, \citenamefont {Scheel},\ and\ \citenamefont
  {Teukolsky}}]{Bhagwat:2017tkm}%
  \BibitemOpen
  \bibfield  {author} {\bibinfo {author} {\bibfnamefont {S.}~\bibnamefont
  {Bhagwat}}, \bibinfo {author} {\bibfnamefont {M.}~\bibnamefont {Okounkova}},
  \bibinfo {author} {\bibfnamefont {S.~W.}\ \bibnamefont {Ballmer}}, \bibinfo
  {author} {\bibfnamefont {D.~A.}\ \bibnamefont {Brown}}, \bibinfo {author}
  {\bibfnamefont {M.}~\bibnamefont {Giesler}}, \bibinfo {author} {\bibfnamefont
  {M.~A.}\ \bibnamefont {Scheel}}, \ and\ \bibinfo {author} {\bibfnamefont
  {S.~A.}\ \bibnamefont {Teukolsky}},\ }\href {\doibase
  10.1103/PhysRevD.97.104065} {\bibfield  {journal} {\bibinfo  {journal} {Phys.
  Rev. D}\ }\textbf {\bibinfo {volume} {97}},\ \bibinfo {pages} {104065}
  (\bibinfo {year} {2018})},\ \Eprint {http://arxiv.org/abs/1711.00926}
  {arXiv:1711.00926 [gr-qc]} \BibitemShut {NoStop}%
\bibitem [{\citenamefont {Cook}(2020)}]{Cook:2020otn}%
  \BibitemOpen
  \bibfield  {author} {\bibinfo {author} {\bibfnamefont {G.~B.}\ \bibnamefont
  {Cook}},\ }\href {\doibase 10.1103/PhysRevD.102.024027} {\bibfield  {journal}
  {\bibinfo  {journal} {Phys. Rev. D}\ }\textbf {\bibinfo {volume} {102}},\
  \bibinfo {pages} {024027} (\bibinfo {year} {2020})},\ \Eprint
  {http://arxiv.org/abs/2004.08347} {arXiv:2004.08347 [gr-qc]} \BibitemShut
  {NoStop}%
\bibitem [{\citenamefont {Mourier}\ \emph {et~al.}(2021)\citenamefont
  {Mourier}, \citenamefont {Jim\'enez~Forteza}, \citenamefont {Pook-Kolb},
  \citenamefont {Krishnan},\ and\ \citenamefont {Schnetter}}]{Mourier:2020mwa}%
  \BibitemOpen
  \bibfield  {author} {\bibinfo {author} {\bibfnamefont {P.}~\bibnamefont
  {Mourier}}, \bibinfo {author} {\bibfnamefont {X.}~\bibnamefont
  {Jim\'enez~Forteza}}, \bibinfo {author} {\bibfnamefont {D.}~\bibnamefont
  {Pook-Kolb}}, \bibinfo {author} {\bibfnamefont {B.}~\bibnamefont {Krishnan}},
  \ and\ \bibinfo {author} {\bibfnamefont {E.}~\bibnamefont {Schnetter}},\
  }\href {\doibase 10.1103/PhysRevD.103.044054} {\bibfield  {journal} {\bibinfo
   {journal} {Phys. Rev. D}\ }\textbf {\bibinfo {volume} {103}},\ \bibinfo
  {pages} {044054} (\bibinfo {year} {2021})},\ \Eprint
  {http://arxiv.org/abs/2010.15186} {arXiv:2010.15186 [gr-qc]} \BibitemShut
  {NoStop}%
\bibitem [{\citenamefont {Finch}\ and\ \citenamefont
  {Moore}(2021)}]{Finch:2021iip}%
  \BibitemOpen
  \bibfield  {author} {\bibinfo {author} {\bibfnamefont {E.}~\bibnamefont
  {Finch}}\ and\ \bibinfo {author} {\bibfnamefont {C.~J.}\ \bibnamefont
  {Moore}},\ }\href {\doibase 10.1103/PhysRevD.103.084048} {\bibfield
  {journal} {\bibinfo  {journal} {Phys. Rev. D}\ }\textbf {\bibinfo {volume}
  {103}},\ \bibinfo {pages} {084048} (\bibinfo {year} {2021})},\ \Eprint
  {http://arxiv.org/abs/2102.07794} {arXiv:2102.07794 [gr-qc]} \BibitemShut
  {NoStop}%
\bibitem [{\citenamefont {Forteza}\ and\ \citenamefont
  {Mourier}(2021)}]{Forteza:2021wfq}%
  \BibitemOpen
  \bibfield  {author} {\bibinfo {author} {\bibfnamefont {X.~J.}\ \bibnamefont
  {Forteza}}\ and\ \bibinfo {author} {\bibfnamefont {P.}~\bibnamefont
  {Mourier}},\ }\href {\doibase 10.1103/PhysRevD.104.124072} {\bibfield
  {journal} {\bibinfo  {journal} {Phys. Rev. D}\ }\textbf {\bibinfo {volume}
  {104}},\ \bibinfo {pages} {124072} (\bibinfo {year} {2021})},\ \Eprint
  {http://arxiv.org/abs/2107.11829} {arXiv:2107.11829 [gr-qc]} \BibitemShut
  {NoStop}%
\bibitem [{\citenamefont {Berti}\ \emph
  {et~al.}(2006{\natexlab{b}})\citenamefont {Berti}, \citenamefont {Cardoso},\
  and\ \citenamefont {Casals}}]{Berti:2005gp}%
  \BibitemOpen
  \bibfield  {author} {\bibinfo {author} {\bibfnamefont {E.}~\bibnamefont
  {Berti}}, \bibinfo {author} {\bibfnamefont {V.}~\bibnamefont {Cardoso}}, \
  and\ \bibinfo {author} {\bibfnamefont {M.}~\bibnamefont {Casals}},\ }\href
  {\doibase 10.1103/PhysRevD.73.109902} {\bibfield  {journal} {\bibinfo
  {journal} {Phys. Rev. D}\ }\textbf {\bibinfo {volume} {73}},\ \bibinfo
  {pages} {024013} (\bibinfo {year} {2006}{\natexlab{b}})},\ \bibinfo {note}
  {[Erratum: Phys.Rev.D 73, 109902 (2006)]},\ \Eprint
  {http://arxiv.org/abs/gr-qc/0511111} {arXiv:gr-qc/0511111} \BibitemShut
  {NoStop}%
\bibitem [{\citenamefont {Teukolsky}(1973)}]{Teukolsky:1973ha}%
  \BibitemOpen
  \bibfield  {author} {\bibinfo {author} {\bibfnamefont {S.~A.}\ \bibnamefont
  {Teukolsky}},\ }\href {\doibase 10.1086/152444} {\bibfield  {journal}
  {\bibinfo  {journal} {Astrophys. J.}\ }\textbf {\bibinfo {volume} {185}},\
  \bibinfo {pages} {635} (\bibinfo {year} {1973})}\BibitemShut {NoStop}%
\bibitem [{\citenamefont {Gleiser}\ \emph {et~al.}(1996)\citenamefont
  {Gleiser}, \citenamefont {Nicasio}, \citenamefont {Price},\ and\
  \citenamefont {Pullin}}]{Gleiser:1995gx}%
  \BibitemOpen
  \bibfield  {author} {\bibinfo {author} {\bibfnamefont {R.~J.}\ \bibnamefont
  {Gleiser}}, \bibinfo {author} {\bibfnamefont {C.~O.}\ \bibnamefont
  {Nicasio}}, \bibinfo {author} {\bibfnamefont {R.~H.}\ \bibnamefont {Price}},
  \ and\ \bibinfo {author} {\bibfnamefont {J.}~\bibnamefont {Pullin}},\ }\href
  {\doibase 10.1088/0264-9381/13/10/001} {\bibfield  {journal} {\bibinfo
  {journal} {Class. Quant. Grav.}\ }\textbf {\bibinfo {volume} {13}},\ \bibinfo
  {pages} {L117} (\bibinfo {year} {1996})},\ \Eprint
  {http://arxiv.org/abs/gr-qc/9510049} {arXiv:gr-qc/9510049} \BibitemShut
  {NoStop}%
\bibitem [{\citenamefont {Brizuela}\ \emph {et~al.}(2009)\citenamefont
  {Brizuela}, \citenamefont {Martin-Garcia},\ and\ \citenamefont
  {Tiglio}}]{Brizuela:2009qd}%
  \BibitemOpen
  \bibfield  {author} {\bibinfo {author} {\bibfnamefont {D.}~\bibnamefont
  {Brizuela}}, \bibinfo {author} {\bibfnamefont {J.~M.}\ \bibnamefont
  {Martin-Garcia}}, \ and\ \bibinfo {author} {\bibfnamefont {M.}~\bibnamefont
  {Tiglio}},\ }\href {\doibase 10.1103/PhysRevD.80.024021} {\bibfield
  {journal} {\bibinfo  {journal} {Phys. Rev. D}\ }\textbf {\bibinfo {volume}
  {80}},\ \bibinfo {pages} {024021} (\bibinfo {year} {2009})},\ \Eprint
  {http://arxiv.org/abs/0903.1134} {arXiv:0903.1134 [gr-qc]} \BibitemShut
  {NoStop}%
\bibitem [{\citenamefont {Ioka}\ and\ \citenamefont
  {Nakano}(2007)}]{Ioka:2007ak}%
  \BibitemOpen
  \bibfield  {author} {\bibinfo {author} {\bibfnamefont {K.}~\bibnamefont
  {Ioka}}\ and\ \bibinfo {author} {\bibfnamefont {H.}~\bibnamefont {Nakano}},\
  }\href {\doibase 10.1103/PhysRevD.76.061503} {\bibfield  {journal} {\bibinfo
  {journal} {Phys. Rev. D}\ }\textbf {\bibinfo {volume} {76}},\ \bibinfo
  {pages} {061503} (\bibinfo {year} {2007})},\ \Eprint
  {http://arxiv.org/abs/0704.3467} {arXiv:0704.3467 [astro-ph]} \BibitemShut
  {NoStop}%
\bibitem [{\citenamefont {Nakano}\ and\ \citenamefont
  {Ioka}(2007)}]{Nakano:2007cj}%
  \BibitemOpen
  \bibfield  {author} {\bibinfo {author} {\bibfnamefont {H.}~\bibnamefont
  {Nakano}}\ and\ \bibinfo {author} {\bibfnamefont {K.}~\bibnamefont {Ioka}},\
  }\href {\doibase 10.1103/PhysRevD.76.084007} {\bibfield  {journal} {\bibinfo
  {journal} {Phys. Rev. D}\ }\textbf {\bibinfo {volume} {76}},\ \bibinfo
  {pages} {084007} (\bibinfo {year} {2007})},\ \Eprint
  {http://arxiv.org/abs/0708.0450} {arXiv:0708.0450 [gr-qc]} \BibitemShut
  {NoStop}%
\bibitem [{\citenamefont {Pazos}\ \emph {et~al.}(2010)\citenamefont {Pazos},
  \citenamefont {Brizuela}, \citenamefont {Martin-Garcia},\ and\ \citenamefont
  {Tiglio}}]{Pazos:2010xf}%
  \BibitemOpen
  \bibfield  {author} {\bibinfo {author} {\bibfnamefont {E.}~\bibnamefont
  {Pazos}}, \bibinfo {author} {\bibfnamefont {D.}~\bibnamefont {Brizuela}},
  \bibinfo {author} {\bibfnamefont {J.~M.}\ \bibnamefont {Martin-Garcia}}, \
  and\ \bibinfo {author} {\bibfnamefont {M.}~\bibnamefont {Tiglio}},\ }\href
  {\doibase 10.1103/PhysRevD.82.104028} {\bibfield  {journal} {\bibinfo
  {journal} {Phys. Rev. D}\ }\textbf {\bibinfo {volume} {82}},\ \bibinfo
  {pages} {104028} (\bibinfo {year} {2010})},\ \Eprint
  {http://arxiv.org/abs/1009.4665} {arXiv:1009.4665 [gr-qc]} \BibitemShut
  {NoStop}%
\bibitem [{\citenamefont {London}\ \emph {et~al.}(2014)\citenamefont {London},
  \citenamefont {Shoemaker},\ and\ \citenamefont {Healy}}]{London:2014cma}%
  \BibitemOpen
  \bibfield  {author} {\bibinfo {author} {\bibfnamefont {L.}~\bibnamefont
  {London}}, \bibinfo {author} {\bibfnamefont {D.}~\bibnamefont {Shoemaker}}, \
  and\ \bibinfo {author} {\bibfnamefont {J.}~\bibnamefont {Healy}},\ }\href
  {\doibase 10.1103/PhysRevD.90.124032} {\bibfield  {journal} {\bibinfo
  {journal} {Phys. Rev. D}\ }\textbf {\bibinfo {volume} {90}},\ \bibinfo
  {pages} {124032} (\bibinfo {year} {2014})},\ \bibinfo {note} {[Erratum:
  Phys.Rev.D 94, 069902 (2016)]},\ \Eprint {http://arxiv.org/abs/1404.3197}
  {arXiv:1404.3197 [gr-qc]} \BibitemShut {NoStop}%
\bibitem [{\citenamefont {Loutrel}\ \emph {et~al.}(2021)\citenamefont
  {Loutrel}, \citenamefont {Ripley}, \citenamefont {Giorgi},\ and\
  \citenamefont {Pretorius}}]{Loutrel:2020wbw}%
  \BibitemOpen
  \bibfield  {author} {\bibinfo {author} {\bibfnamefont {N.}~\bibnamefont
  {Loutrel}}, \bibinfo {author} {\bibfnamefont {J.~L.}\ \bibnamefont {Ripley}},
  \bibinfo {author} {\bibfnamefont {E.}~\bibnamefont {Giorgi}}, \ and\ \bibinfo
  {author} {\bibfnamefont {F.}~\bibnamefont {Pretorius}},\ }\href {\doibase
  10.1103/PhysRevD.103.104017} {\bibfield  {journal} {\bibinfo  {journal}
  {Phys. Rev. D}\ }\textbf {\bibinfo {volume} {103}},\ \bibinfo {pages}
  {104017} (\bibinfo {year} {2021})},\ \Eprint
  {http://arxiv.org/abs/2008.11770} {arXiv:2008.11770 [gr-qc]} \BibitemShut
  {NoStop}%
\bibitem [{\citenamefont {Ripley}\ \emph {et~al.}(2021)\citenamefont {Ripley},
  \citenamefont {Loutrel}, \citenamefont {Giorgi},\ and\ \citenamefont
  {Pretorius}}]{Ripley:2020xby}%
  \BibitemOpen
  \bibfield  {author} {\bibinfo {author} {\bibfnamefont {J.~L.}\ \bibnamefont
  {Ripley}}, \bibinfo {author} {\bibfnamefont {N.}~\bibnamefont {Loutrel}},
  \bibinfo {author} {\bibfnamefont {E.}~\bibnamefont {Giorgi}}, \ and\ \bibinfo
  {author} {\bibfnamefont {F.}~\bibnamefont {Pretorius}},\ }\href {\doibase
  10.1103/PhysRevD.103.104018} {\bibfield  {journal} {\bibinfo  {journal}
  {Phys. Rev. D}\ }\textbf {\bibinfo {volume} {103}},\ \bibinfo {pages}
  {104018} (\bibinfo {year} {2021})},\ \Eprint
  {http://arxiv.org/abs/2010.00162} {arXiv:2010.00162 [gr-qc]} \BibitemShut
  {NoStop}%
\bibitem [{\citenamefont {Berti}\ and\ \citenamefont
  {Klein}(2014)}]{Berti:2014fga}%
  \BibitemOpen
  \bibfield  {author} {\bibinfo {author} {\bibfnamefont {E.}~\bibnamefont
  {Berti}}\ and\ \bibinfo {author} {\bibfnamefont {A.}~\bibnamefont {Klein}},\
  }\href {\doibase 10.1103/PhysRevD.90.064012} {\bibfield  {journal} {\bibinfo
  {journal} {Phys. Rev. D}\ }\textbf {\bibinfo {volume} {90}},\ \bibinfo
  {pages} {064012} (\bibinfo {year} {2014})},\ \Eprint
  {http://arxiv.org/abs/1408.1860} {arXiv:1408.1860 [gr-qc]} \BibitemShut
  {NoStop}%
\bibitem [{\citenamefont {Maga\~na Zertuche}\ \emph {et~al.}(2022)\citenamefont
  {Maga\~na Zertuche} \emph {et~al.}}]{MaganaZertuche:2021syq}%
  \BibitemOpen
  \bibfield  {author} {\bibinfo {author} {\bibfnamefont {L.}~\bibnamefont
  {Maga\~na Zertuche}} \emph {et~al.},\ }\href {\doibase
  10.1103/PhysRevD.105.104015} {\bibfield  {journal} {\bibinfo  {journal}
  {Phys. Rev. D}\ }\textbf {\bibinfo {volume} {105}},\ \bibinfo {pages}
  {104015} (\bibinfo {year} {2022})},\ \Eprint
  {http://arxiv.org/abs/2110.15922} {arXiv:2110.15922 [gr-qc]} \BibitemShut
  {NoStop}%
\bibitem [{\citenamefont {Sberna}\ \emph {et~al.}(2022)\citenamefont {Sberna},
  \citenamefont {Bosch}, \citenamefont {East}, \citenamefont {Green},\ and\
  \citenamefont {Lehner}}]{Sberna:2021eui}%
  \BibitemOpen
  \bibfield  {author} {\bibinfo {author} {\bibfnamefont {L.}~\bibnamefont
  {Sberna}}, \bibinfo {author} {\bibfnamefont {P.}~\bibnamefont {Bosch}},
  \bibinfo {author} {\bibfnamefont {W.~E.}\ \bibnamefont {East}}, \bibinfo
  {author} {\bibfnamefont {S.~R.}\ \bibnamefont {Green}}, \ and\ \bibinfo
  {author} {\bibfnamefont {L.}~\bibnamefont {Lehner}},\ }\href {\doibase
  10.1103/PhysRevD.105.064046} {\bibfield  {journal} {\bibinfo  {journal}
  {Phys. Rev. D}\ }\textbf {\bibinfo {volume} {105}},\ \bibinfo {pages}
  {064046} (\bibinfo {year} {2022})},\ \Eprint
  {http://arxiv.org/abs/2112.11168} {arXiv:2112.11168 [gr-qc]} \BibitemShut
  {NoStop}%
\bibitem [{\citenamefont {Ma}\ \emph {et~al.}(2022)\citenamefont {Ma},
  \citenamefont {Mitman}, \citenamefont {Sun}, \citenamefont {Deppe},
  \citenamefont {H\'ebert}, \citenamefont {Kidder}, \citenamefont {Moxon},
  \citenamefont {Throwe}, \citenamefont {Vu},\ and\ \citenamefont
  {Chen}}]{Ma:2022wpv}%
  \BibitemOpen
  \bibfield  {author} {\bibinfo {author} {\bibfnamefont {S.}~\bibnamefont
  {Ma}}, \bibinfo {author} {\bibfnamefont {K.}~\bibnamefont {Mitman}}, \bibinfo
  {author} {\bibfnamefont {L.}~\bibnamefont {Sun}}, \bibinfo {author}
  {\bibfnamefont {N.}~\bibnamefont {Deppe}}, \bibinfo {author} {\bibfnamefont
  {F.}~\bibnamefont {H\'ebert}}, \bibinfo {author} {\bibfnamefont {L.~E.}\
  \bibnamefont {Kidder}}, \bibinfo {author} {\bibfnamefont {J.}~\bibnamefont
  {Moxon}}, \bibinfo {author} {\bibfnamefont {W.}~\bibnamefont {Throwe}},
  \bibinfo {author} {\bibfnamefont {N.~L.}\ \bibnamefont {Vu}}, \ and\ \bibinfo
  {author} {\bibfnamefont {Y.}~\bibnamefont {Chen}},\ }\href@noop {} {\
  (\bibinfo {year} {2022})},\ \Eprint {http://arxiv.org/abs/2207.10870}
  {arXiv:2207.10870 [gr-qc]} \BibitemShut {NoStop}%
\bibitem [{\citenamefont {Sperhake}\ \emph {et~al.}(2008)\citenamefont
  {Sperhake}, \citenamefont {Cardoso}, \citenamefont {Pretorius}, \citenamefont
  {Berti},\ and\ \citenamefont {Gonzalez}}]{Sperhake:2008ga}%
  \BibitemOpen
  \bibfield  {author} {\bibinfo {author} {\bibfnamefont {U.}~\bibnamefont
  {Sperhake}}, \bibinfo {author} {\bibfnamefont {V.}~\bibnamefont {Cardoso}},
  \bibinfo {author} {\bibfnamefont {F.}~\bibnamefont {Pretorius}}, \bibinfo
  {author} {\bibfnamefont {E.}~\bibnamefont {Berti}}, \ and\ \bibinfo {author}
  {\bibfnamefont {J.~A.}\ \bibnamefont {Gonzalez}},\ }\href {\doibase
  10.1103/PhysRevLett.101.161101} {\bibfield  {journal} {\bibinfo  {journal}
  {Phys. Rev. Lett.}\ }\textbf {\bibinfo {volume} {101}},\ \bibinfo {pages}
  {161101} (\bibinfo {year} {2008})},\ \Eprint {http://arxiv.org/abs/0806.1738}
  {arXiv:0806.1738 [gr-qc]} \BibitemShut {NoStop}%
\bibitem [{\citenamefont {Healy}\ \emph {et~al.}(2016)\citenamefont {Healy},
  \citenamefont {Ruchlin}, \citenamefont {Lousto},\ and\ \citenamefont
  {Zlochower}}]{Healy:2015mla}%
  \BibitemOpen
  \bibfield  {author} {\bibinfo {author} {\bibfnamefont {J.}~\bibnamefont
  {Healy}}, \bibinfo {author} {\bibfnamefont {I.}~\bibnamefont {Ruchlin}},
  \bibinfo {author} {\bibfnamefont {C.~O.}\ \bibnamefont {Lousto}}, \ and\
  \bibinfo {author} {\bibfnamefont {Y.}~\bibnamefont {Zlochower}},\ }\href
  {\doibase 10.1103/PhysRevD.94.104020} {\bibfield  {journal} {\bibinfo
  {journal} {Phys. Rev. D}\ }\textbf {\bibinfo {volume} {94}},\ \bibinfo
  {pages} {104020} (\bibinfo {year} {2016})},\ \Eprint
  {http://arxiv.org/abs/1506.06153} {arXiv:1506.06153 [gr-qc]} \BibitemShut
  {NoStop}%
\bibitem [{\citenamefont {Clough}\ \emph {et~al.}(2015)\citenamefont {Clough},
  \citenamefont {Figueras}, \citenamefont {Finkel}, \citenamefont {Kunesch},
  \citenamefont {Lim},\ and\ \citenamefont {Tunyasuvunakool}}]{Clough:2015sqa}%
  \BibitemOpen
  \bibfield  {author} {\bibinfo {author} {\bibfnamefont {K.}~\bibnamefont
  {Clough}}, \bibinfo {author} {\bibfnamefont {P.}~\bibnamefont {Figueras}},
  \bibinfo {author} {\bibfnamefont {H.}~\bibnamefont {Finkel}}, \bibinfo
  {author} {\bibfnamefont {M.}~\bibnamefont {Kunesch}}, \bibinfo {author}
  {\bibfnamefont {E.~A.}\ \bibnamefont {Lim}}, \ and\ \bibinfo {author}
  {\bibfnamefont {S.}~\bibnamefont {Tunyasuvunakool}},\ }\href {\doibase
  10.1088/0264-9381/32/24/245011} {\bibfield  {journal} {\bibinfo  {journal}
  {Class. Quant. Grav.}\ }\textbf {\bibinfo {volume} {32}},\ \bibinfo {pages}
  {245011} (\bibinfo {year} {2015})},\ \Eprint
  {http://arxiv.org/abs/1503.03436} {arXiv:1503.03436 [gr-qc]} \BibitemShut
  {NoStop}%
\bibitem [{\citenamefont {Andrade}\ \emph {et~al.}(2021)\citenamefont {Andrade}
  \emph {et~al.}}]{Andrade:2021rbd}%
  \BibitemOpen
  \bibfield  {author} {\bibinfo {author} {\bibfnamefont {T.}~\bibnamefont
  {Andrade}} \emph {et~al.},\ }\href {\doibase 10.21105/joss.03703} {\bibfield
  {journal} {\bibinfo  {journal} {J. Open Source Softw.}\ }\textbf {\bibinfo
  {volume} {6}},\ \bibinfo {pages} {3703} (\bibinfo {year} {2021})},\ \Eprint
  {http://arxiv.org/abs/2201.03458} {arXiv:2201.03458 [gr-qc]} \BibitemShut
  {NoStop}%
\bibitem [{\citenamefont {Pretorius}(2005)}]{Pretorius:2004jg}%
  \BibitemOpen
  \bibfield  {author} {\bibinfo {author} {\bibfnamefont {F.}~\bibnamefont
  {Pretorius}},\ }\href {\doibase 10.1088/0264-9381/22/2/014} {\bibfield
  {journal} {\bibinfo  {journal} {Class. Quant. Grav.}\ }\textbf {\bibinfo
  {volume} {22}},\ \bibinfo {pages} {425} (\bibinfo {year} {2005})},\ \Eprint
  {http://arxiv.org/abs/gr-qc/0407110} {arXiv:gr-qc/0407110} \BibitemShut
  {NoStop}%
\bibitem [{\citenamefont {Cook}\ \emph {et~al.}(2016)\citenamefont {Cook},
  \citenamefont {Figueras}, \citenamefont {Kunesch}, \citenamefont {Sperhake},\
  and\ \citenamefont {Tunyasuvunakool}}]{Cook:2016soy}%
  \BibitemOpen
  \bibfield  {author} {\bibinfo {author} {\bibfnamefont {W.~G.}\ \bibnamefont
  {Cook}}, \bibinfo {author} {\bibfnamefont {P.}~\bibnamefont {Figueras}},
  \bibinfo {author} {\bibfnamefont {M.}~\bibnamefont {Kunesch}}, \bibinfo
  {author} {\bibfnamefont {U.}~\bibnamefont {Sperhake}}, \ and\ \bibinfo
  {author} {\bibfnamefont {S.}~\bibnamefont {Tunyasuvunakool}},\ }\href
  {\doibase 10.1142/S0218271816410133} {\bibfield  {journal} {\bibinfo
  {journal} {Int. J. Mod. Phys. D}\ }\textbf {\bibinfo {volume} {25}},\
  \bibinfo {pages} {1641013} (\bibinfo {year} {2016})},\ \Eprint
  {http://arxiv.org/abs/1603.00362} {arXiv:1603.00362 [gr-qc]} \BibitemShut
  {NoStop}%
\bibitem [{\citenamefont {Shibata}\ and\ \citenamefont
  {Yoshino}(2010)}]{Shibata:2010wz}%
  \BibitemOpen
  \bibfield  {author} {\bibinfo {author} {\bibfnamefont {M.}~\bibnamefont
  {Shibata}}\ and\ \bibinfo {author} {\bibfnamefont {H.}~\bibnamefont
  {Yoshino}},\ }\href {\doibase 10.1103/PhysRevD.81.104035} {\bibfield
  {journal} {\bibinfo  {journal} {Phys. Rev. D}\ }\textbf {\bibinfo {volume}
  {81}},\ \bibinfo {pages} {104035} (\bibinfo {year} {2010})},\ \Eprint
  {http://arxiv.org/abs/1004.4970} {arXiv:1004.4970 [gr-qc]} \BibitemShut
  {NoStop}%
\bibitem [{\citenamefont {Helfer}\ \emph {et~al.}()\citenamefont {Helfer} \emph
  {et~al.}}]{Headonpaper}%
  \BibitemOpen
  \bibfield  {author} {\bibinfo {author} {\bibfnamefont {T.}~\bibnamefont
  {Helfer}} \emph {et~al.},\ }\href@noop {} {\ }\bibinfo {note} {In
  preparation}\BibitemShut {NoStop}%
\bibitem [{\citenamefont {Boyle}\ \emph {et~al.}(2019)\citenamefont {Boyle}
  \emph {et~al.}}]{Boyle:2019kee}%
  \BibitemOpen
  \bibfield  {author} {\bibinfo {author} {\bibfnamefont {M.}~\bibnamefont
  {Boyle}} \emph {et~al.},\ }\href {\doibase 10.1088/1361-6382/ab34e2}
  {\bibfield  {journal} {\bibinfo  {journal} {Class. Quant. Grav.}\ }\textbf
  {\bibinfo {volume} {36}},\ \bibinfo {pages} {195006} (\bibinfo {year}
  {2019})},\ \Eprint {http://arxiv.org/abs/1904.04831} {arXiv:1904.04831
  [gr-qc]} \BibitemShut {NoStop}%
\bibitem [{\citenamefont {Moxon}\ \emph {et~al.}(2020)\citenamefont {Moxon},
  \citenamefont {Scheel},\ and\ \citenamefont {Teukolsky}}]{Moxon:2020gha}%
  \BibitemOpen
  \bibfield  {author} {\bibinfo {author} {\bibfnamefont {J.}~\bibnamefont
  {Moxon}}, \bibinfo {author} {\bibfnamefont {M.~A.}\ \bibnamefont {Scheel}}, \
  and\ \bibinfo {author} {\bibfnamefont {S.~A.}\ \bibnamefont {Teukolsky}},\
  }\href {\doibase 10.1103/PhysRevD.102.044052} {\bibfield  {journal} {\bibinfo
   {journal} {Phys. Rev. D}\ }\textbf {\bibinfo {volume} {102}},\ \bibinfo
  {pages} {044052} (\bibinfo {year} {2020})},\ \Eprint
  {http://arxiv.org/abs/2007.01339} {arXiv:2007.01339 [gr-qc]} \BibitemShut
  {NoStop}%
\bibitem [{\citenamefont {Moxon}\ \emph {et~al.}(2021)\citenamefont {Moxon},
  \citenamefont {Scheel}, \citenamefont {Teukolsky}, \citenamefont {Deppe},
  \citenamefont {Fischer}, \citenamefont {H\'ebert}, \citenamefont {Kidder},\
  and\ \citenamefont {Throwe}}]{Moxon:2021gbv}%
  \BibitemOpen
  \bibfield  {author} {\bibinfo {author} {\bibfnamefont {J.}~\bibnamefont
  {Moxon}}, \bibinfo {author} {\bibfnamefont {M.~A.}\ \bibnamefont {Scheel}},
  \bibinfo {author} {\bibfnamefont {S.~A.}\ \bibnamefont {Teukolsky}}, \bibinfo
  {author} {\bibfnamefont {N.}~\bibnamefont {Deppe}}, \bibinfo {author}
  {\bibfnamefont {N.}~\bibnamefont {Fischer}}, \bibinfo {author} {\bibfnamefont
  {F.}~\bibnamefont {H\'ebert}}, \bibinfo {author} {\bibfnamefont {L.~E.}\
  \bibnamefont {Kidder}}, \ and\ \bibinfo {author} {\bibfnamefont
  {W.}~\bibnamefont {Throwe}},\ }\href@noop {} {\  (\bibinfo {year} {2021})},\
  \Eprint {http://arxiv.org/abs/2110.08635} {arXiv:2110.08635 [gr-qc]}
  \BibitemShut {NoStop}%
\bibitem [{\citenamefont {Mitman}\ \emph {et~al.}(2023)\citenamefont {Mitman}
  \emph {et~al.}}]{Mitman:2022qdl}%
  \BibitemOpen
  \bibfield  {author} {\bibinfo {author} {\bibfnamefont {K.}~\bibnamefont
  {Mitman}} \emph {et~al.},\ }\href {\doibase 10.1103/PhysRevLett.130.081402}
  {\bibfield  {journal} {\bibinfo  {journal} {Phys. Rev. Lett.}\ }\textbf
  {\bibinfo {volume} {130}},\ \bibinfo {pages} {081402} (\bibinfo {year}
  {2023})},\ \Eprint {http://arxiv.org/abs/2208.07380} {arXiv:2208.07380
  [gr-qc]} \BibitemShut {NoStop}%
\bibitem [{\citenamefont {Stein}(2019)}]{Stein:2019mop}%
  \BibitemOpen
  \bibfield  {author} {\bibinfo {author} {\bibfnamefont {L.~C.}\ \bibnamefont
  {Stein}},\ }\href {\doibase 10.21105/joss.01683} {\bibfield  {journal}
  {\bibinfo  {journal} {J. Open Source Softw.}\ }\textbf {\bibinfo {volume}
  {4}},\ \bibinfo {pages} {1683} (\bibinfo {year} {2019})},\ \Eprint
  {http://arxiv.org/abs/1908.10377} {arXiv:1908.10377 [gr-qc]} \BibitemShut
  {NoStop}%
\bibitem [{\citenamefont {Hunter}(2007)}]{matplotlib}%
  \BibitemOpen
  \bibfield  {author} {\bibinfo {author} {\bibfnamefont {J.~D.}\ \bibnamefont
  {Hunter}},\ }\href {\doibase 10.1109/MCSE.2007.55} {\bibfield  {journal}
  {\bibinfo  {journal} {Comput. Sci. Eng.}\ }\textbf {\bibinfo {volume} {9}},\
  \bibinfo {pages} {90} (\bibinfo {year} {2007})}\BibitemShut {NoStop}%
\bibitem [{\citenamefont {Harris}\ \emph {et~al.}(2020)\citenamefont {Harris}
  \emph {et~al.}}]{numpy}%
  \BibitemOpen
  \bibfield  {author} {\bibinfo {author} {\bibfnamefont {C.~R.}\ \bibnamefont
  {Harris}} \emph {et~al.},\ }\href {\doibase 10.1038/s41586-020-2649-2}
  {\bibfield  {journal} {\bibinfo  {journal} {Nature (London)}\ }\textbf
  {\bibinfo {volume} {585}},\ \bibinfo {pages} {357} (\bibinfo {year}
  {2020})}\BibitemShut {NoStop}%
\bibitem [{\citenamefont {Stanzione}\ \emph {et~al.}(2020)\citenamefont
  {Stanzione}, \citenamefont {West}, \citenamefont {Evans}, \citenamefont
  {Minyard}, \citenamefont {Ghattas},\ and\ \citenamefont
  {Panda}}]{10.1145/3311790.3396656}%
  \BibitemOpen
  \bibfield  {author} {\bibinfo {author} {\bibfnamefont {D.}~\bibnamefont
  {Stanzione}}, \bibinfo {author} {\bibfnamefont {J.}~\bibnamefont {West}},
  \bibinfo {author} {\bibfnamefont {R.~T.}\ \bibnamefont {Evans}}, \bibinfo
  {author} {\bibfnamefont {T.}~\bibnamefont {Minyard}}, \bibinfo {author}
  {\bibfnamefont {O.}~\bibnamefont {Ghattas}}, \ and\ \bibinfo {author}
  {\bibfnamefont {D.~K.}\ \bibnamefont {Panda}},\ }in\ \href {\doibase
  10.1145/3311790.3396656} {\emph {\bibinfo {booktitle} {Practice and
  Experience in Advanced Research Computing}}},\ \bibinfo {series and number}
  {PEARC '20}\ (\bibinfo  {publisher} {Association for Computing Machinery},\
  \bibinfo {address} {New York, NY, USA},\ \bibinfo {year} {2020})\ p.\
  \bibinfo {pages} {106–111}\BibitemShut {NoStop}%
\end{thebibliography}%

\clearpage

\appendix
\section*{Supplemental material}

\begin{figure*}
	\includegraphics[width=\textwidth]{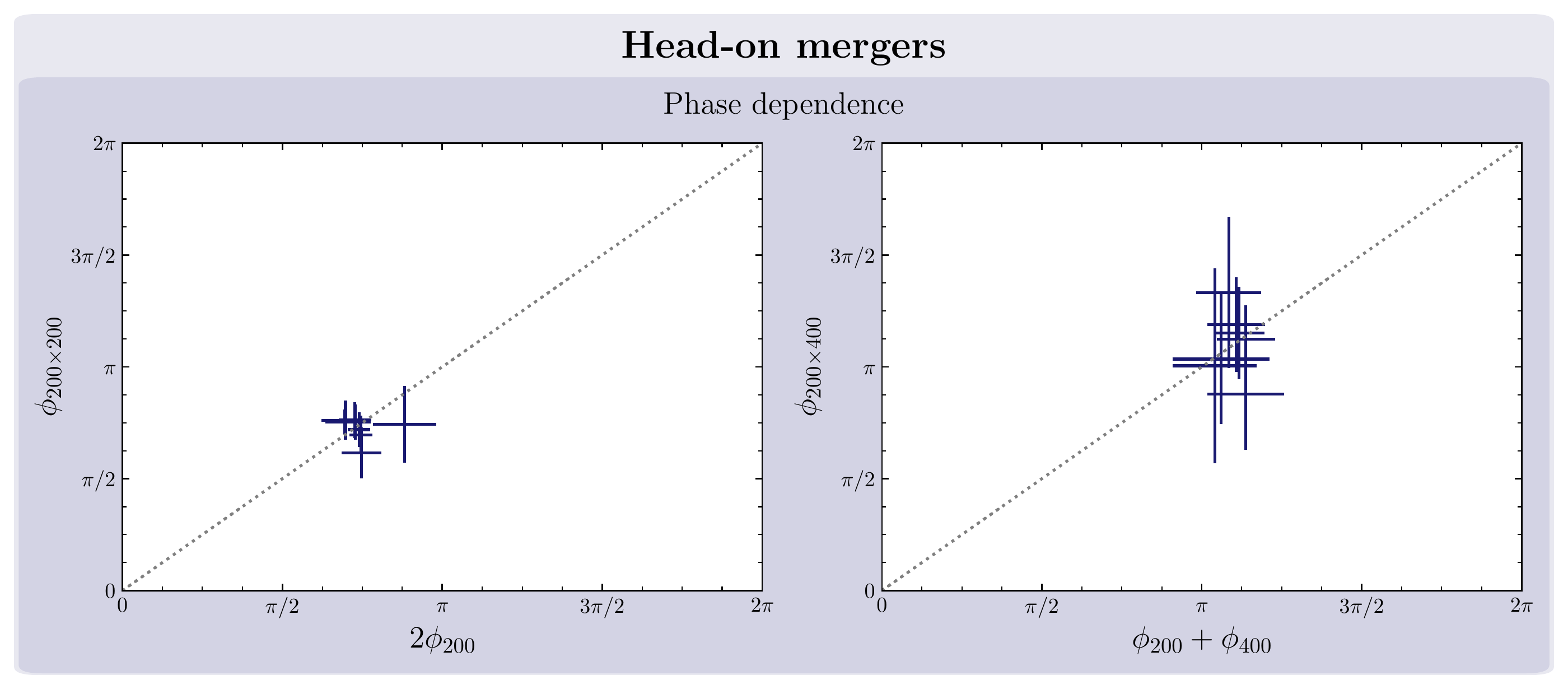}
	\caption{\label{fig:headon_phase} 
      The relationship between the phase of the second-order modes and that of the sourcing first-order modes for the head-on merger simulations, similar to the right column of Fig.~\ref{fig:amplitude_dependence}.
      Different crosses correspond to simulations with different boosts $\gamma$, and the width and height of the crosses are the errors.
      The gray dotted line is the expected relationship.}
\end{figure*}

\noindent
{\bf \em Fitting coefficients for amplitude and phase.}
If we assume that Eq.~\eqref{eq:dependence} in the main text holds, we can fit the data points shown in Figs.~\ref{fig:amplitude_dependence} and \ref{fig:headon_phase} for the coefficient of proportionality of Eq.~\eqref{eq:amplitude_dependence} and the constant in Eq.~\eqref{eq:phase_dependence}.

We find
\begin{align*}
    A_{200 \times 200} &= (1.700 \pm 0.115) A_{200}^2, \\
    A_{200 \times 400} &= (2.868 \pm 0.295) A_{200}A_{400}, \\ 
    \phi_{200 \times 200} &= 2 \phi_{200} + (4.422 \pm 0.116), \\
    \phi_{200 \times 400} &= \phi_{200} + \phi_{400} + (4.384 \pm 0.304), 
\end{align*}
for the head-on mergers, and
\begin{align*}
    A_{220 \times 220} &= (0.1637 \pm 0.0018) A_{220}^2, \\
    A_{220 \times 330} &= (0.4735 \pm 0.0062) A_{220}A_{330},\\
    \phi_{220 \times 220} &= 2 \phi_{220} + (0.177 \pm 0.036), \\
    \phi_{220 \times 330} &= \phi_{220} + \phi_{330} + (0.143 \pm 0.028), 
\end{align*}
for the quasicircular mergers.
Note that, in general, the coefficients may depend on the remnant BH spin.
For quasicircular mergers, our fit of the $220 \times 220$ mode coefficient in the amplitude dependence equation is in excellent agreement with Ref.~\cite{Mitman:2022qdl}. The coefficients of the $220 \times 220$ and $220 \times 330$ mode amplitudes are also in good order-of-magnitude agreement with the estimates in Ref.~\cite{London:2014cma}.

\noindent
{\bf \em Head-on merger simulations.}
We use the \textsc{GRChombo} codebase to simulate the head-on merger of two BHs with Lorentz boost factor $\gamma$ (see~\cite{Headonpaper} for details). The axial symmetry along the collision axis allows us to use dimensional reduction (see \cite{Cook:2016soy,Pretorius:2004jg,Shibata:2010wz}) and simulate the system using only two dimensions, significantly decreasing the computational cost and improving the precision of the simulations. We use initial data similar to Ref.~\cite{Healy:2015mla}, consisting of a superposition of two boosted BHs. The constraint violations in the initial data are mitigated by using a large initial distance $R_{\rm init}$ between the two BHs (typically $R_{\rm init} = 1600 - 2000$ $M_0$, where $M_0$ is the rest mass of the BHs).

\begin{table*}[t]
    \centering
    \renewcommand{\arraystretch}{1.5}
    \begin{tabular}{ | c | c c c c c c c c|}
    \hline
    Simulation & $q$ & $\chi_{1,z}$ & $\chi_{2,z}$ & $\chi_{\rm eff}$ & $M$ & $|\chi|$ & $220\times220$ & $220\times330$ \\
    \hline
    SXS:BBH:1155 & 1.00 & 0.00 & 0.00 & 0.00 & 0.9516 & 0.6864 & \checkmark & \\
    SXS:BBH:1513 & 1.15 & -0.11 & -0.01 & -0.06 & 0.9537 & 0.6636 & \checkmark & \\
    SXS:BBH:0312 & 1.20 & 0.39 & -0.48 & 0.00 & 0.9515 & 0.6972 & \checkmark & \\
    SXS:BBH:0305 & 1.22 & 0.33 & -0.44 & -0.02 & 0.9520 & 0.6921 & \checkmark & \\
    SXS:BBH:1143 & 1.25 & 0.00 & 0.00 & 0.00 & 0.9528 & 0.6795 & \checkmark & \checkmark\\
    SXS:BBH:1511 & 1.47 & 0.03 & -0.10 & -0.02 & 0.9552 & 0.6637 & \checkmark & \checkmark\\
    SXS:BBH:0593 & 1.50 & 0.00 & 0.00 & 0.00 & 0.9552 & 0.6641 & \checkmark & \checkmark\\
    SXS:BBH:0547 & 1.70 & 0.21 & -0.34 & 0.01 & 0.9561 & 0.6820 &  & \checkmark\\
    SXS:BBH:1354 & 1.83 & 0.00 & 0.00 & 0.00 & 0.9592 & 0.6377 &  & \checkmark\\
    SXS:BBH:0403 & 1.88 & 0.00 & -0.05 & -0.02 & 0.9601 & 0.6303 &  & \checkmark\\
    SXS:BBH:1365 & 2.00 & 0.00 & 0.00 & 0.00 & 0.9613 & 0.6230 &  & \checkmark\\
    \hline
    \end{tabular}
    \caption{SXS waveform catalog used for our fits.
    The $x$ and $y$ components of both progenitor spins are zero for all simulations.}
    \label{tab:SXS_simulations}
\end{table*}

\begin{table*}[t]
    \centering
    \renewcommand{\arraystretch}{1.5}
    \begin{tabular}{ | c | c | c | c | c | c |}
    \hline
    \multicolumn{3}{|c|}{Head-on mergers} & \multicolumn{3}{c|}{Quasicircular mergers} \\
    \hline
    Simulation & $t_{{\rm opt}, 40}/M$ & $t_{{\rm opt}, 60}/M$ & Simulation & $t_{{\rm opt}, 44}/M$ & $t_{{\rm opt}, 55}/M$\\
    \hline
    $\gamma = 1.20$ & $5.1$ & $7.7$& SXS:BBH:1155 & $27.9$ & \\
    $\gamma = 1.30$ & $10.3$& $17.5$& SXS:BBH:1513 & $28.6$ & \\
    $\gamma = 1.40$ & $12.0$& $5.0$& SXS:BBH:0312 & $25.0$ & \\
    $\gamma = 1.50$ & $11.3$& $10.4$& SXS:BBH:0305 & $25.9$ & \\
    $\gamma = 1.75$ & $17.4$& $18.1$& SXS:BBH:1143 & $27.0$ & $29.9$ \\
    $\gamma = 2.00$ & $16.2$& $18.5$& SXS:BBH:1511 & $30.0$ & $31.3$ \\
    $\gamma = 2.30$ & $16.9$& $23.1$& SXS:BBH:0593 & 30.5 & $31.1$\\
    $\gamma = 3.20$ & $15.4$& & SXS:BBH:0547 & & $31.6$\\
     & & & SXS:BBH:1354 & & 30.4\\
     & & & SXS:BBH:0403 & & 30.9\\
     & & & SXS:BBH:1365 & & 30.0\\
    \hline
    \end{tabular}
    \caption{Optimal fit starting time that gives a frequency closest to the expected quadratic mode of the $\ell m$ harmonic, $t_{{\rm opt}, \ell m}/M$, for head-on simulations with different $\gamma$ factors (left) and for quasicircular mergers from the SXS catalog (right).}
    \label{tab:t_opt}
\end{table*}

\noindent
{\bf \em Selection of SXS waveforms.}
We use a conservative criterion to select the SXS simulations included in our study:
(i) we require the free-mode search to yield a fitted frequency that is closer to the nonlinear mode than any other possible linear mode for a duration of $T_0$;
(ii) when we include a second-order mode with fixed frequency in our fitting model, we require the fitted amplitude to be consistent within $10 \%$ in a fitting window of length $T_0$ around the time where the fitted frequency in the free mode search is closest to the expected one.
We find that only nonprecessing systems with individual progenitor dimensionless spins $\chi_1$, $\chi_2 \lesssim 0.5$ meet these criteria. %
For the $220 \times 220$ mode search, we only use simulations with mass ratio $q \leq 1.50$, while for the $220 \times 330$ mode search we only use those with $1.25 \leq q \leq 2.00$, because simulations outside of these ranges do not meet the above criteria.
We include all simulations that meet the criteria, excluding those with almost overlapping simulation parameters.
We find the nonlinear modes consistently for most of these simulations, with the exception of earlier simulations that do not give accurate higher $\ell m$ multipoles (e.g., SXS:BBH:0194), or simulations where the higher $\ell m$ multipoles do not settle down to a clean QNM decay before $t - t_0 \sim 35 M$ (e.g., SXS:BBH:0605).

The simulation numbers used and their corresponding parameters are shown in Table~\ref{tab:SXS_simulations}.
We stress again that the selection criteria are conservative, and if we allow for a larger variation of the amplitude we could include additional simulations that are still consistent with the quadratic amplitude dependence, albeit with larger error bars. 
We expect the quality of the fits to improve for waveforms using Cauchy Characteristic Extraction~\cite{Moxon:2020gha,Moxon:2021gbv,Mitman:2022qdl}, which are not yet publicly available.

\begin{figure*}[t]
	\includegraphics[width=\textwidth]{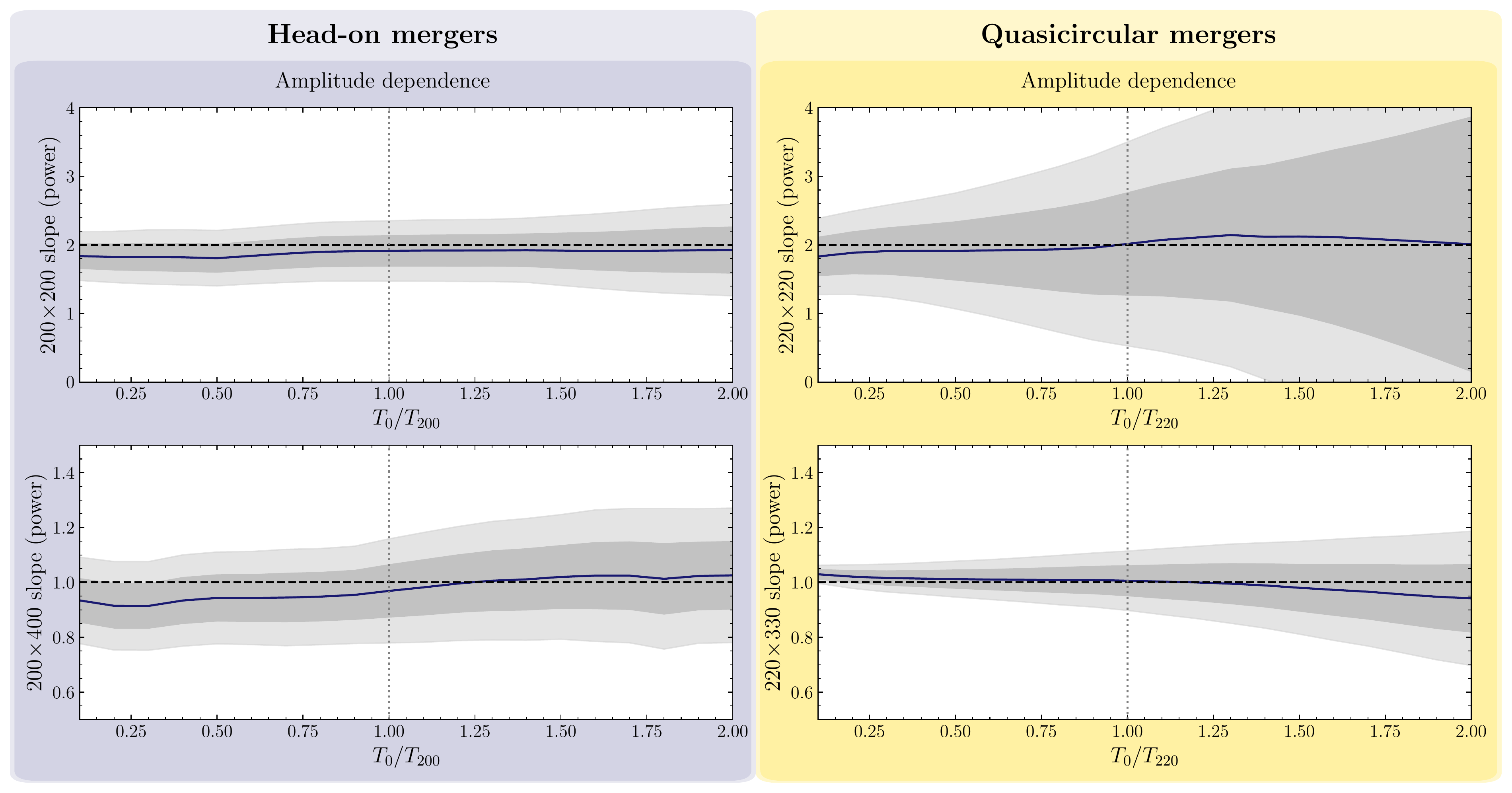}
	\caption{\label{fig:slope_var} 
      Variation of the slope of the amplitude dependence plots (left and middle columns of Fig.~\ref{fig:amplitude_dependence}) when we change the length of the starting time window $T_0$ used for extracting the amplitude.
      The horizontal black dashed line is the expected slope.
      The dark gray and light gray bands label the $1 \sigma$ and $2 \sigma$ confidence intervals, respectively.
      The vertical gray dashed line corresponds to the value of $T_0$ used in the main text.}
\end{figure*}

\noindent
{\bf \em Optimal starting times of the fits.}
In the main text, we checked the consistency of our fits around an optimal starting time of the fits, defined as the starting time $t_{{\rm opt}, \ell m}$ that gives a fitted frequency closest to the expected quadratic mode in the $\ell m$ harmonic. For reproducibility, in Table~\ref{tab:t_opt} we list $t_{{\rm opt}, \ell m}/M$ for each of the head-on and quasicircular merger simulations we considered.

\noindent
{\bf \em Variation of the starting time window.}
The estimated total error on the amplitudes often has a significant contribution from the amplitude fluctuation error $\delta A_{\rm  fluc}$. This is computed as follows: we first find the fit starting time that gives a frequency closest to the expected one in a free mode search, then we compute the standard deviation of the amplitude within a time window of length $T_{0}$ around this time.
In Fig.~\ref{fig:slope_var} we show that while varying the length of the window affects the errors in the inferred slope of the amplitude dependence, the results are still consistent with prediction within $1\sigma$ unless the time window is very narrow, i.e. $T_0 \lesssim 0.5 T_{200 / 220}$.

\begin{figure*}
	\includegraphics[width=\textwidth]{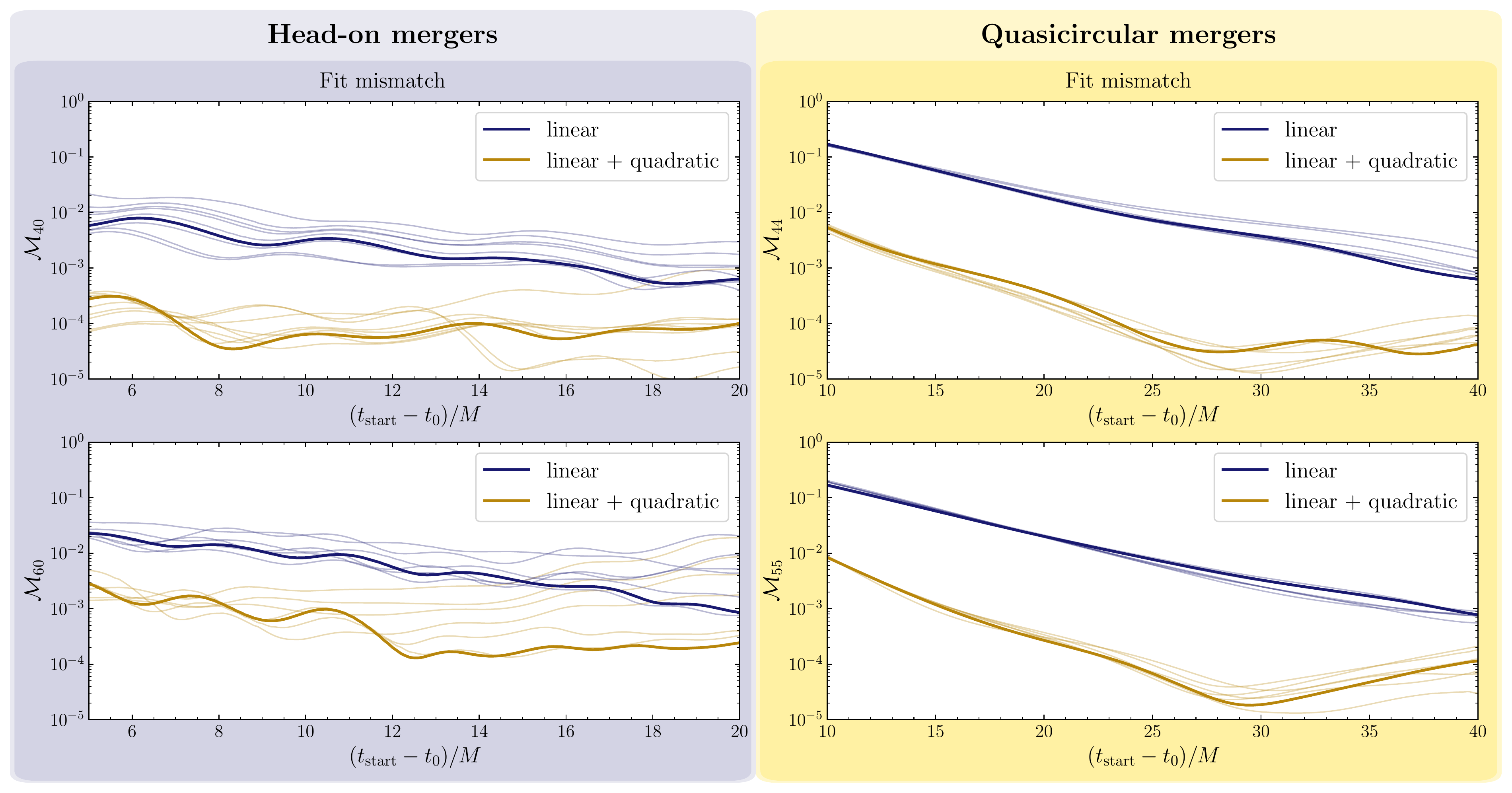}
	\caption{\label{fig:mismatch} 
      The mismatch $\mathcal{M}_{\ell m}$ between the fitted waveform and the actual waveform in the $\ell m$ harmonic of the head-on (left column) and quasi-circular merger (right column) simulations.
      The blue curves corresponds to fits consisting containing only the linear modes identified in the waveform, while the gold curves include also the corresponding quadratic mode.
      The highlighted lines correspond to $\gamma = 2$ for the head-on mergers, and SXS:BBH:0305 (top) and SXS:BBH:0403 (bottom) for the quasicircular mergers.
      For each individual simulation, the model containing the quadratic mode always has smaller mismatch.
      }
\end{figure*}

\noindent
{\bf \em Waveform mismatch between fitted and actual waveforms.}
If a quadratic mode is present in the waveform, a model containing such a mode should fit the waveform significantly better than one that only includes linear modes.
The ``goodness of fit'' can be quantified by considering the mismatch between the best fit waveform $\psi$ and the actual waveform $\Psi$,
\begin{equation}
\mathcal{M} = 1 - \dfrac{\langle \Psi | \psi \rangle}{\sqrt{\langle \Psi | \Psi \rangle \langle \psi | \psi \rangle}},
\end{equation}
where the scalar product is defined as
\begin{equation}
    \langle f | g \rangle = \int^{t_{\rm end}}_{t_{\rm start}} f(t) g^*(t) dt,
    \label{eq:mismatch}
\end{equation}
the asterisk denotes complex conjugation, and $t_{\rm end}$ is the ending time of the fitting window.
In Fig.~\ref{fig:mismatch} we show that the mismatch always decreases significantly when the model includes a quadratic mode, rather than only the other (linear) modes.

\noindent
{\bf \em Estimate of errors on the mode amplitudes.}
We estimate the errors on the amplitudes, which include contributions from the fluctuation of the amplitude within the $T_{0}$ window, the error between simulation resolutions, and the waveform extraction errors, as follows.

For the SXS simulations, we take into account the fluctuation (standard deviation) $\delta A_{\rm fluc}$ of the amplitude within the $T_{0}$ window, the difference $\delta A_{\rm res}$ of the amplitude extracted from the highest and second-highest resolutions of the SXS simulations, and the difference $\delta A_{\rm extrap}$ of the amplitude extracted from the waveforms with extrapolation orders of $N = 2$ and $N=3$.
The total error is estimated as $\delta A_{\rm tot} = \sqrt{\delta A_{\rm fluc}^2 + \delta A_{\rm res}^2 + \delta A_{\rm extrap}^2}$.

For the head-on merger simulations, we estimate the error as $\delta A_{\rm tot} = \sqrt{\delta A_{\rm fluc}^2 + \delta A_{\rm res}^2 + \delta A_{\rm extract}^2 + \delta A_{\rm mass}^2}$, where $\delta A_{\rm res}$ is computed by Richardson extrapolation, $\delta A_{\rm extract}$ is the variation of the amplitude when the waveform is extracted at the radii $900, 1000, \dots 1500 M_0$, where $M_0$ is the total rest mass of the binary, and $\delta A_{\rm mass}$ is propagated from the error of the estimated remnant mass.

We do not include $\delta A_{\rm mass}$ for the SXS waveforms because this contribution is negligible.

\noindent
{\bf \em Second-order phase dependence for head-on mergers.}
In the right column of Fig.~\ref{fig:amplitude_dependence} of the main text we show that the extracted phases of the SXS simulations follow the expected relationship, Eq.~\eqref{eq:phase_dependence} of the main text.
Figure~\ref{fig:headon_phase} shows analogous plots for the $200 \times 200$ and $200 \times 400$ modes of the head-on mergers. The results are again consistent with the expected relationship, although the extracted phases are less accurate when compared to the SXS results.
This is likely because the head-on simulation waveforms are strictly real (while the quasicircular SXS waveforms are complex), which makes it harder to extract the phase evolution of the waveform.

\begin{figure*}
	\includegraphics[width=\textwidth]{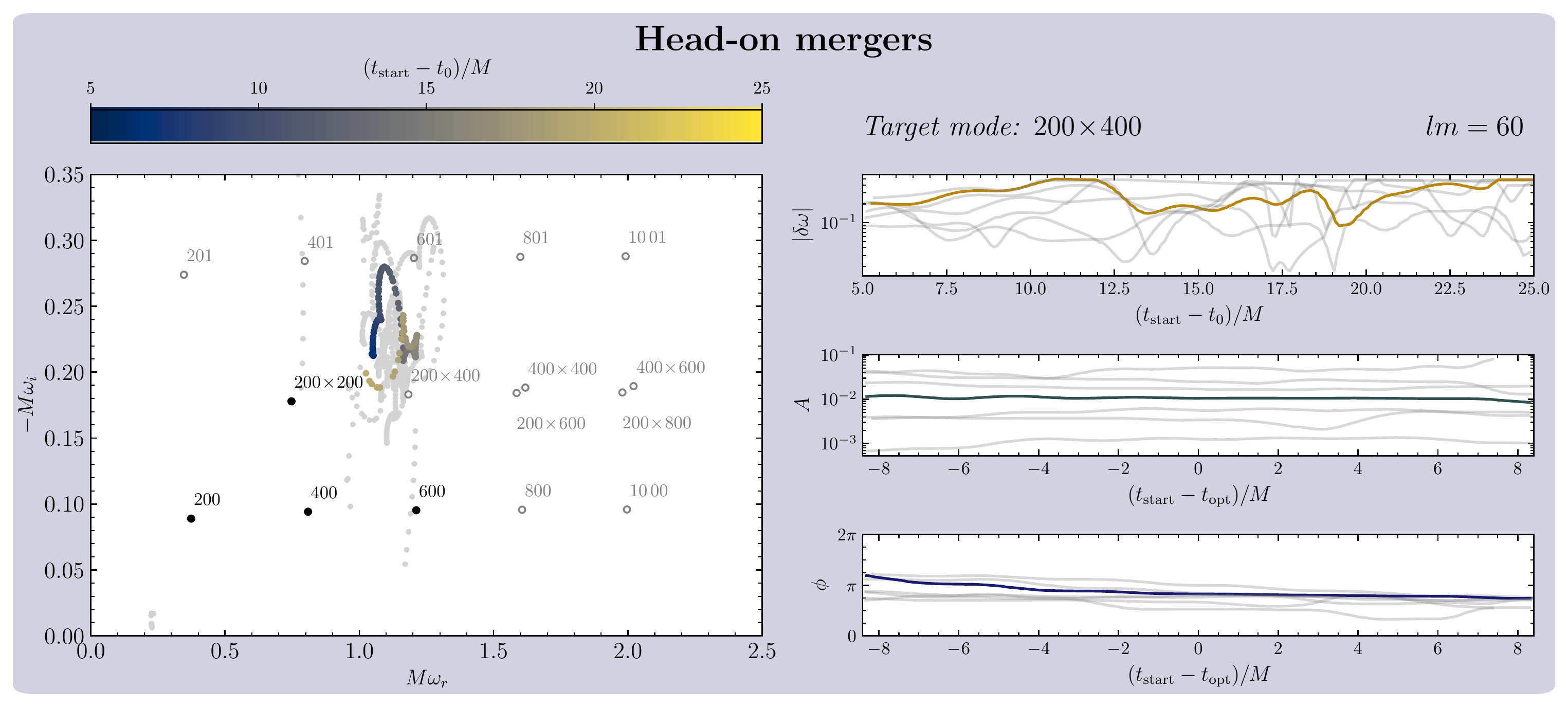}
	\caption{\label{fig:200x400} 
      Results for the $200 \times 400$ mode search in the $\ell m = 60$ harmonic for head-on mergers (see Fig.~\ref{fig:mainplot}).
      Free mode search (left); deviation of fitted frequency from expected second-order value (top right); amplitude (center right) and phase (bottom right) of the fitted mode around the starting time with lowest $|\delta \omega|$, with frequency fixed at the second-order value.}
\end{figure*}

\noindent
{\bf \em The $200 \times 400$ mode of head-on mergers.}
As discussed in the main text, we find strong evidence for second-order modes. In particular the $\ell m=55$ multipole contains the $220\times 330$ mode, which is sourced by two different first-order modes. 
By the same argument, we also expect to observe the $200 \times 400$ mode in the $\ell m=60$ multipole of head-on merger waveforms.
As shown in Fig.~\ref{fig:200x400}, the free mode search for the $200 \times 400$ mode does not converge as well as that for the $200 \times 200$ mode, while the real parts of the fitted frequencies are arguably consistent with the expected mode.
The extracted amplitudes and phases are also not as flat as those of the $200 \times 200$ mode, even varying by more than $10 \%$ within the $T_0$ window for some values of $\gamma$.
Note that similar to linear modes like the $200$ and $400$ modes, the $200 \times 200$ mode is mixed into the $\ell m = 60$ harmonic due to numerical contamination, and we have to include it in our fitting model along with the other linear modes (shown as black circles in Fig.~\ref{fig:200x400}).
While higher-resolution simulations might be required to cleanly extract the $200 \times 400$ mode, if we assume that the $200 \times 400$ mode exists and extract its amplitude and phase, we find that the values are still consistent with Eqs.~\eqref{eq:amplitude_dependence} and~\eqref{eq:phase_dependence} of the main text, as shown in the bottom left panel of Fig.~\ref{eq:amplitude_dependence} and in the right panel of Fig.~\ref{fig:headon_phase}.

\begin{figure*}
	\includegraphics[width=\textwidth]{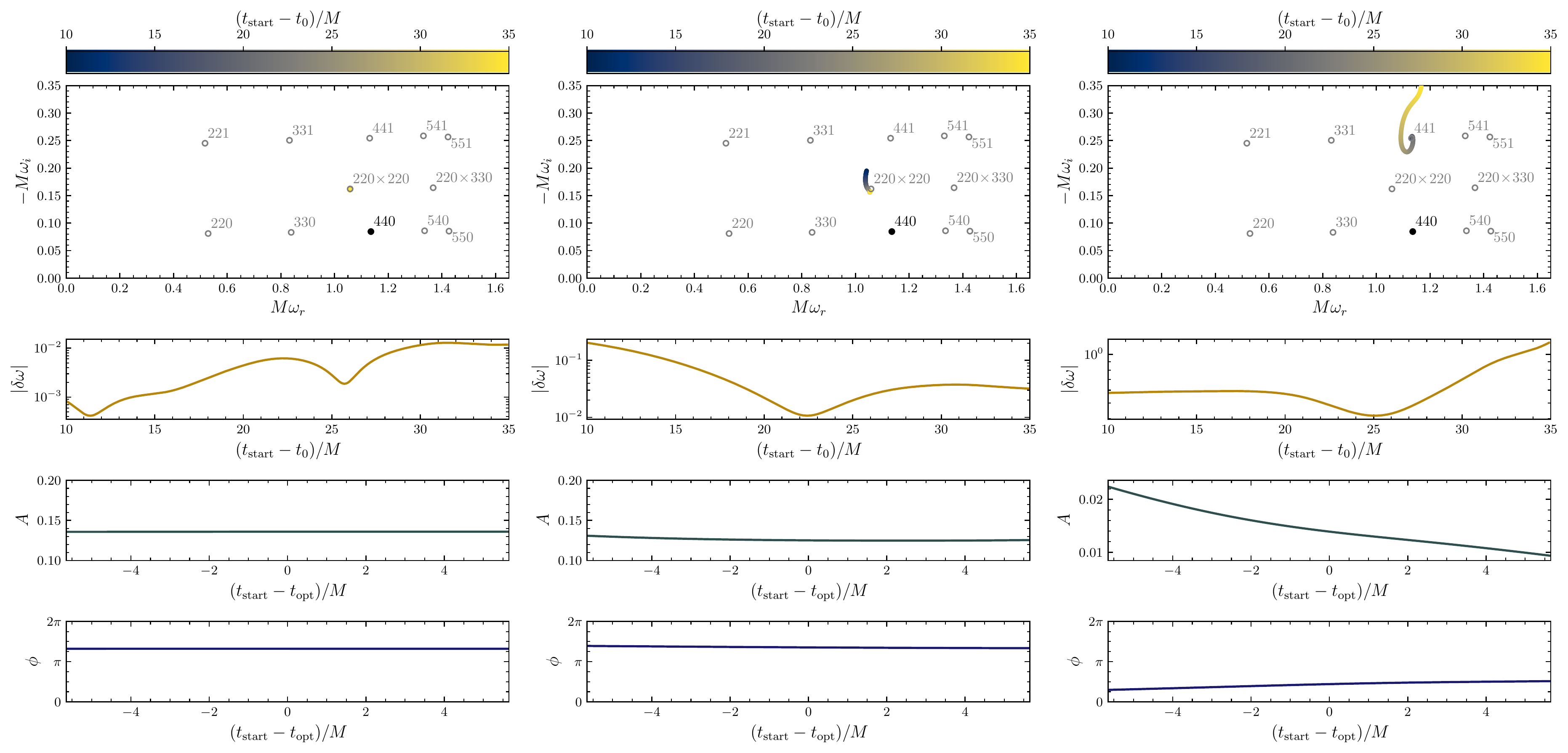}
	\caption{\label{fig:toymodel} 
      Results for toy model A (left column), B (middle column) and C (right column).}
\end{figure*}

\noindent
{\bf \em Controlled experiments to test mode-fitting criteria.}
A good way to test our mode searching procedure is by constructing toy model waveforms with simulated noise that we can control.
We consider three simulated scenarios: the first (toy model A) is a contamination-free idealized scenario; the second (toy model B) is the realistic scenario we expect to encounter in our simulations; the third (toy model C) serves to show the distinguishability of nonlinear modes from linear modes with similar frequencies.

We first test the ability of our procedure to identify modes in waveforms with noise.
We construct toy model A with two injected modes plus numerical noise to mimic the $\ell m = 44$ harmonic waveform of the SXS:BBH:0305 simulation:
\begin{multline}
    \Psi_A(t) = A_{440} e^{-i (\omega_{440} t + \phi_{440})} \\
            + A_{220\times 220} e^{-i (\omega_{220\times 220} t + \phi_{220\times 220})} + n_{0305, 44}
\end{multline}
The numerical noise in the $\ell m = 44$ SXS:BBH:0305 waveform, $n_{0305, 44}$, is conservatively (over)estimated by computing the difference between the waveform with the highest resolution (Lev 6) with extrapolation order $N = 2$, and the one with second-highest resolution (Lev 5) with $N = 3$.
Before taking the difference, we take care of the cumulative phase error of the waveforms by aligning them such that the mismatch in the ringdown phase is minimum.
The amplitudes and phases of the injected $440$ and $220 \times 220$ modes are taken from the best-fit values at the optimal starting time (i.e., when $|\delta \omega|$ is a minimum in the center column of Fig.~\ref{fig:mainplot}), so as to mimic SXS:BBH:0305 as close as possible.
As seen in the left column of Fig.~\ref{fig:toymodel}, when we fit the waveform with one mode fixed at the $440$ mode frequency and another mode with free frequency, the free mode converges to the $220 \times 220$ frequency, as expected.
Moreover, when we also fix the frequency of the $220 \times 220$ mode, the fitted amplitude and phase are consistent within $10 \%$ in the $T_{0}$ starting time window around the optimal fitting time.

Of course, the waveform is not only contaminated by noise, but also by other modes and by nonlinearities close to the peak of the waveform.
To determine how this contamination would affect our results, we construct toy model B by adding the $441$ overtone:
\begin{equation}
    \Psi_B(t) = \Psi_A(t) + A_{441} e^{-i (\omega_{441} t + \phi_{441})},
\end{equation}
where we take $A_{441} = 5 A_{440}$ and $\phi_{441} = \phi_{440} + \pi$.
We use this value for the phase because the first overtone has been found (at least in the $\ell m = 22$ waveform) to be almost in anti-phase with the fundamental mode, but changing the phase does not affect our results to a significant degree.

As shown in the middle column of Fig.~\ref{fig:toymodel}, the free mode hovers in the vicinity of the $220 \times 220$ frequency, while the amplitude of the fixed-mode fit varies by $\sim 10\%$ across the $T_{0}$ time window.
This shows that if other contaminations are present, even if they decay faster than the $220 \times 220$ mode, we should not expect the free mode search to converge perfectly, nor the fitted amplitude to be perfectly consistent.
Incidentally, we could not confidently resolve the $441$ overtone in our waveform, so we expect that there should be more contaminations (other than the $441$ mode) that are masking it. Therefore the fits in Fig.~\ref{fig:mainplot} of the main text should perform at most as well (if not worse) than the results shown here for toy model B.

Finally, suppose that the $220 \times 220$ mode does not exist in the signal. Would our procedure return false positive results?
This is a particularly important test for the following reason:
it is well known that it is easier to accurately identify the real part of the frequency of a mode in a ringdown waveform than its imaginary part.
By coincidence, the real parts of $\omega_{441}$ and $\omega_{220 \times 220}$ are similar, so we need to make sure that what we are seeing is not the $441$ mode, rather than the $220 \times 220$ mode.
We do this by considering toy model C, where the waveform only consists of the $440$ and $441$ modes:
\begin{multline}
     \Psi_C(t) = A_{440} e^{-i (\omega_{440} t + \phi_{440})} \\
            + A_{441} e^{-i (\omega_{441} t + \phi_{441})} + n_{0305, 44}.
\end{multline}
The amplitude and phase of the $441$ mode are the same as in toy model B.
As shown in the right column of Fig.~\ref{fig:toymodel}, the fitted frequencies for the free-mode search scatter around $\omega_{441}$, significantly further away from $\omega_{220\times 220}$ than the results in Fig.~\ref{fig:mainplot}.
The fitted amplitude also varies by a factor of $\sim 1$, much more than $\sim 10 \%$.
However, note that while the $220\times 220$ mode is not present, the fitted phase varies by less than $\sim 10 \% \times 2 \pi$, because the $220 \times 220$ and $441$ modes have similar real frequencies.
This shows that it is important to consider the consistency of both the amplitude and the phase when searching for the second-order modes, which often have a real frequency close to other linear modes.

In conclusion, toy models A, B, and C confirm that our mode-searching procedure is robust against numerical noise and against contamination by faster-decaying modes.
The criterion of an amplitude consistent within $\sim 10 \%$ in a time window of $T_{0}$ is also adequate to rule out false positives, while accounting for the variation caused by contamination.

\end{document}